\documentclass[notitlepage,nofootinbib,preprintnumbers,aps,prd, superscriptaddress]{revtex4-1}
\usepackage{amsmath,amssymb,natbib,bm,color}
\usepackage{tikz}
\usepackage{appendix}
\usepackage{float}
\usepackage[caption = false]{subfig}

\begin{document}

\title{Asymmetric Reheating by Primordial Black Holes}
\author{Pearl Sandick}
\affiliation{Department of Physics and Astronomy, University of Utah, Salt Lake City, UT 84112, USA}
\author{Barmak Shams Es Haghi}
\affiliation{Department of Physics and Astronomy, University of Utah, Salt Lake City, UT 84112, USA}
\author{Kuver Sinha}
\affiliation{Department of Physics and Astronomy, University of Oklahoma, Norman, OK 73019, USA}

\begin{abstract}

We investigate Hawking evaporation of a population of primordial black holes (PBHs) prior to Big Bang Nucleosynthesis (BBN) as a mechanism to achieve asymmetric reheating of two sectors coupled solely by gravity. While the visible sector is reheated by the inflaton or a modulus, the dark sector is reheated by PBHs. Compared to inflationary or modular  reheating of both sectors, there are two advantages: $(i)$ inflaton or moduli mediated operators that can subsequently thermalize the dark sector with the visible sector are not relevant to the asymmetric reheating process; $(ii)$ the mass and abundance of the PBHs provide parametric control of the thermal history of the dark sector, and in particular the ratio of the temperatures of the two sectors. Asymmetric reheating with PBHs turns out to have a particularly rich dark sector phenomenology, which we explore using a single self-interacting real scalar field in the dark sector as a template. Four thermal histories, involving non-relativistic and relativistic dark matter (DM) at chemical equilibrium, followed by the presence or absence of cannibalism, are explored. These histories are then constrained by the observed relic abundance in the current Universe and the Bullet Cluster. The case where PBHs dominate the energy density of the Universe, and reheat both the visible as well as the dark sectors, is also treated in detail.

\end{abstract}

\maketitle

\section{Introduction}

The identity of dark matter (DM) is one of the main open problems in particle physics and cosmology, and remains  elusive in spite of numerous direct and indirect detection experiments, as well as collider searches. Since the evidence for the existence of DM comes from its gravitational interaction with ordinary matter, scenarios in which DM interacts \textit{only} gravitationally with the visible sector are well-motivated. Indeed, this minimal description of DM does not exclude the possibility of a rich, equilibrated dark sector, with the observed relic abundance being set by self-interactions among dark sector fields \footnote{Self interacting DM has been introduced to resolve tensions between small-scale structure observations and N-body simulations of collisionless cold DM. Problems such as the ``cusp vs core problem''~\cite{Flores:1994gz, Moore:1994yx, Oh:2010mc} and the ``too-big-to-fail problem''~\cite{BoylanKolchin:2011de, BoylanKolchin:2011dk, Garrison-Kimmel:2014vqa, Papastergis:2014aba} can possibly be explained by a sizable self-interaction among DM particles~\cite{BoylanKolchin:2011de, Spergel:1999mh, deBlok:2009sp}.}.

An outstanding question for a dark sector that interacts only gravitationally with the visible sector is: how was such a putative sector populated in the first place? This question was originally asked and answered in the context of ``mirror models"  \cite{Berezhiani:1995am, Hodges:1993yb}, and revisited more recently by \cite{Adshead:2016xxj, Hardy:2017wkr, Adshead:2019uwj}.  The same mechanism - inflationary reheating 
 - that populates the visible sector in the early Universe is assumed to also be responsible for populating the dark sector \footnote{Of course, if one relaxes the condition that the dark sector interact only gravitationally, a plethora of mechanisms become available by coupling it to the visible sector through various portals of different strengths~\cite{Feng:2008mu, Ackerman:2008kmp, Kaplan:2009de, Fan:2013yva, Foot:2014uba, Pospelov:2007mp, Pappadopulo:2016pkp, Berlin:2016vnh, Faraggi:2000pv, Chu:2011be, Bernal:2015xba, Heikinheimo:2016yds, Batell:2017cmf, Cheung:2010gj, Berlin:2017ftj, Fitzpatrick:2020vba}. Depending on the strength of the interaction, dark sectors both in thermal equilibrium with as well as thermally decoupled from the visible sector are possible.}. The produced particles in each sector are initially far from equilibrium, but elastic scattering processes and number changing processes can drive each sector towards equilibrium. Although this happens quickly in the visible sector (due to the gauge structure and soft scattering processes)~\cite{Enqvist:1990dp, Chung:1998rq, McDonald:1999hd, Davidson:2000er}, the thermal fate of the dark sector depends on its internal interactions and the initial number density of its particles. The competition between the expansion rate of the Universe and the  rate of self-interactions of DM particles determines the extent of the equilibration in the dark sector, which ranges from not reaching equilibrium at all to kinetic or chemical equilibrium. 

Reheating the dark sector using the inflaton or a modulus comes with certain challenges, as pointed out by~\cite{Adshead:2016xxj, Adshead:2019uwj}. Any asymmetry in the temperatures of the two sectors should be protected against washout due to equilibration by exchanging heavy modes in the spectrum, unavoidably the inflation or modulus itself. In fact, such equilibration occurs quite generically in parameter space, and avoiding this can impose restrictions on the inflaton mass and coupling, or require a non-standard cosmology. Dark sectors with dark radiation that were in thermal equilibrium with the visible sector at any point in their history will be increasingly tightly constrained by future bounds on the effective number of neutrino species, $\Delta N_{\rm eff}$, making this an important issue for model-building in such cases. 

The purpose of this paper is to study a novel reheating mechanism which can reheat two sectors asymmetrically: the Hawking evaporation of a population of primordial black holes (PBHs) with masses in the $0.1\,\text{g}\lesssim M_\text{BH}\lesssim 10^9\,\text{g}$ range. 
Since PBHs in this mass range evaporate before Big Bang Nucleosynthesis (BBN), their abundance is not constrained (they may be accessible to future gravitational wave experiments~\cite{Inomata:2020lmk, Domenech:2020ssp, Papanikolaou:2020qtd}). We will be agnostic about the origin of the PBHs, assuming they exist at some stage in the post-inflationary universe \footnote{The formation of light PBHs in the early universe has been studied using various methods, such as the collapse from inhomogeneities~\cite{Carr:1975qj, Nadezhin:1978}, sudden reduction in the pressure~\cite{Khlopov:1980mg, Khlopov:1981mg, Khlopov:1982mg}, collapse of cosmic loops~\cite{Hawking:1987bn, Polnarev:1988dh, Hansen:1999su, Hogan:1984zb, Nagasawa:2005hv}, bubble collisions~\cite{Crawford:1982yz, Hawking:1982ga, La:1989st, Moss:1994iq, Khlopov:1998nm, Konoplich:1999qq}, and collapse of domain walls~\cite{Rubin:2000dq, Rubin:2001yw, Dokuchaev:2004kr}. For a recent review of PBH formation, we refer to ~\cite{Carr:2020gox} and references therein.}. The Hawking evaporation of PBHs has been extensively studied as a potential explanation for baryogenesis, DM, dark radiation, and axion-like particles~\cite{Hawking:1974rv, Zeldovich:1976vw, Hook:2014mla, Hamada:2016jnq, Hooper:2020otu, Perez-Gonzalez:2020vnz, Datta:2020bht, Baumann:2007yr, Fujita:2014hha, Morrison:2018xla, Chaudhuri:2020wjo, JyotiDas:2021shi, Lennon:2017tqq, Allahverdi:2017sks, Hooper:2019gtx, Masina:2020xhk, Baldes:2020nuv, Gondolo:2020uqv, Bernal:2020kse, Bernal:2020bjf, Bernal:2020ili, Kitabayashi:2021hox, Auffinger:2020afu, Cheek:2021odj, Cheek:2021cfe, Kitabayashi:2021fqx, Hooper:2020evu, Masina:2021zpu, Arbey:2021ysg, Schiavone:2021imu, Bernal:2021yyb}. The idea of thermalization and equilibration of DM particles produced by Hawking evaporation of PBHs by introducing self-interaction has also been studied recently in Ref.~\cite{Bernal:2020kse} to relax bounds from structure formation on light DM.

While asymmetric reheating  is typically susceptible to equilibration due to the presence of operators inducing inflaton or moduli exchange between dark and visible particles, reheating with PBHs is immune to such processes. Since there is no mediator between the two sectors other than  gravity, after equilibrium is established in each sector, any temperature asymmetry  persists and evolves to keep the entropy of each sector conserved during the expansion of the Universe. Moreover, PBH-induced asymmetric reheating exhibits a rich phenomenology that connects the subsequent thermal history of the dark sector with the abundance and mass of PBHs that reheated it. In particular, the initial temperature asymmetry between the visible and dark sectors, which is typically taken as an initial condition in studies of the thermal evolution of dark sectors, now has an origin and can be ``derived" from the properties of the PBH population. 

We explore this rich phenomenology in this paper, taking as a simple template, a minimal self-interacting dark sector consisting of a single real scalar field, which is therefore also the DM particle. The self-interaction strength is assumed to be $\mathcal{O}(1)$ and below the perturbative unitarity limit. The thermal history of the dark sector is best classified in terms of the initial abundance of the DM, which in turn is set by the initial abundance of the PBHs. Four histories are possible as the initial PBH abundance is increased and are schematically depicted in Figure \ref{fig:fig1}. For low abundances, the DM is non-relativistic when it achieves chemical equilibrium and subsequently through its entire history, though it $(i)$ may not (lowest abundance) or $(ii)$ may (higher abundance) enter a cannibal phase after chemical equilibrium and before freeze-out.  For higher abundances, the DM is relativistic when it achieves chemical equilibrium, but again, depending on how high the abundance is, $(iii)$ may not or $(iv)$ may go through a cannibal phase before it freezes out and becomes non-relativistic. 

These four thermal histories are depicted on the parameter space of our scenario, which is comprised of the DM mass, the PBH mass, and the PBH abundance. The results are shown in the left panels of Figures \ref{fig:EQregions} and \ref{fig:abundance}. For two benchmark DM masses (10 MeV and 10 GeV) these thermal histories are depicted on the plane of the relic density versus PBH mass in the left panel of Figure \ref{fig:temps}. The ratio $\xi$ of the dark sector temperature to the visible sector temperature at chemical equilibrium is calculated for the various thermal histories and shown in the  right panels of Figure \ref{fig:temps}. We find that temperature asymmetries $\xi$ larger than, smaller than, and equal to one are all possible. If the dark sector is relativistic at chemical equilibrium, it is always colder than the visible sector at that time. If the dark sector is non-relativistic at chemical equilibrium, it can be hotter or colder than the visible sector.

Reheating the dark sector with PBHs turns out to be a quite predictive framework. Firstly, when  the  initial  temperature  of  PBHs  is smaller than the mass of the DM, equilibrium cannot be established, even when the DM self-coupling saturates the perturbative  unitarity limit. Secondly, we find that self-consistency conditions require that if DM was relativistic at chemical equilibrium, it must undergo a subsequent cannibal phase prior to freeze-out, thereby ruling out one of the four possible thermal histories described above. Furthermore, when subjected to  two conditions: satisfying the relic density in the current Universe, and satisfying bounds from the Bullet Cluster\footnote{The mass range of PBHs we consider already takes care of Cosmic Microwave Background (CMB) and BBN constraints.}, the thermal history of the dark sector becomes even more predictive. DM that is relativistic at chemical equilibrium and then undergoes a cannibal phase before freezing out is found to be incompatible with constraints from Bullet Cluster observations and is thus also ruled out for all DM masses. DM  is therefore forced to be non-relativistic at chemical equilibrium in these scenarios. Moreover, we find that for DM with a mass in the $8\,\text{MeV}\lesssim m_\chi\lesssim 360\,\text{MeV}$ range, the DM could undergo a cannibal phase if the PBH mass is in the $0.1\,\text{g}\lesssim M_\text{BH}\lesssim 205\,\text{g}$ range; otherwise a cannibal phase becomes impossible. These properties are displayed in the right panels of Figure \ref{fig:temps} and show an interconnection between the properties of the PBHs responsible for reheating the dark sector, the DM mass, and observational constraints.

Alongside these results, in every case we show the corresponding results when the initial abundance of the PBHs is large enough to trigger an early matter (PBH)-dominated era. It turns out that in this case, a cannibal phase overproduces DM, regardless of whether DM was relativistic or non-relativistic at chemical equilibrium, thus constraining two of the four possible histories. Non-relativistic chemical equilibrium without a subsequent cannibal phase can give rise to the right relic abundance today for a range of DM and PBH masses. In this case the temperature asymmetry $\xi$ is always larger than one.

The outline of this paper is as follows. In Section~\ref{sec:reheatingbyPBHs}, first we review different possible thermal histories of a self-interacting dark sector populated in the early Universe via some unspecified mechanism by relativistic and far from equilibrium DM particles. We then explore the viable thermal histories of a dark sector populated by Hawking evaporation of PBHs in early Universe. In Section~\ref{sec:results}, after describing the available parameter space and relevant constraints, we present our results. Three appendices are also added: Appendix~\ref{sec:DSgeneral} provides a general setup to review possible thermal histories of a self-interacting dark sector which includes a single scalar field and is populated by an unspecified mechanism in early Universe. In Appendix~\ref{sec:PBH}, we review the formation and Hawking evaporation of PBHs in the early Universe, including the spectra of emitted particles and the condition for transitioning to an early PBH-dominated era. Appendix~\ref{sec:DS} uses the results of Appendices~\ref{sec:DSgeneral} and~\ref{sec:PBH} to provide relevant formulae related to populating a self-interacting dark sector by Hawking evaporation of PBHs and its subsequent equilibration and thermal history.

\section{Reheating two Gravitationally coupled sectors by PBHs}
\label{sec:reheatingbyPBHs}
In this section, we briefly describe different possible thermal histories of a self-interacting dark sector populated in the early Universe. We then explain different outcomes of populating a dark sector by Hawking evaporation of PBHs and the effect of thermalization and equilibration on the DM relic abundance today. 

\subsection{Thermal history of a self-interacting dark sector}
\label{sec:thermalhistory}
A dark sector that is initially populated by relativistic particles far from equilibrium can be driven toward chemical equilibrium if there is a sizeable self-interaction.
The first step towards establishing chemical equilibrium is kinetic equilibrium (thermalization) which requires that the rate of elastic scattering processes, $2\leftrightarrow 2$, becomes comparable to the Hubble expansion rate.

After kinetic equilibrium, the dark sector can generally be described by a temperature and a non-zero chemical potential. Chemical equilibrium requires the chemical potential to become zero, which can occur if $2\leftrightarrow n$, $n>2$, number-changing processes happen fast enough compared to the Hubble expansion rate, and before the particles become non-relativistic.
When the dark sector becomes non-relativistic, $2\rightarrow n$ processes are kinematically forbidden, but $n\rightarrow 2$ processes are still allowed. At this time, if the rate of $n\rightarrow 2$ number-changing processes is less than the Hubble expansion rate, then the dark sector particles decouple and freeze out as cold DM.
On the other hand, if the rate of $n\rightarrow 2$ number-changing processes is comparable to the Hubble expansion rate, the dark sector stays in chemical equilibrium even after it becomes non-relativistic; then $n\rightarrow 2$ processes convert the mass of the DM particles into kinetic energy which slows down the cooling of the temperature of the dark sector due to the expansion and gives rise to the so-called cannibal phase~\cite{Carlson:1992fn}. The cannibal phase will eventually stop when the cold DM particles decouple and freeze out.

\begin{figure}
\subfloat[Non-relativistic at $t_\text{chem-eq}$ and no cannibalism (NRNC)]{\includegraphics[width = 0.8\textwidth]{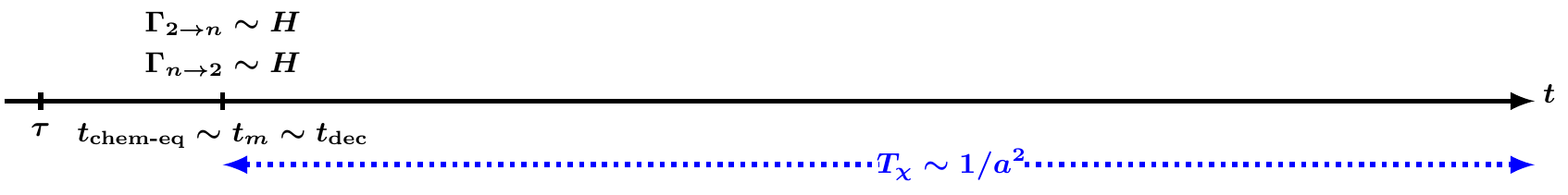}\label{fig:fig1a}} \\
\vspace{3mm}
\subfloat[Non-relativistic at $t_\text{chem-eq}$ and cannibalism (NRC)]{\includegraphics[width = 0.8\textwidth]{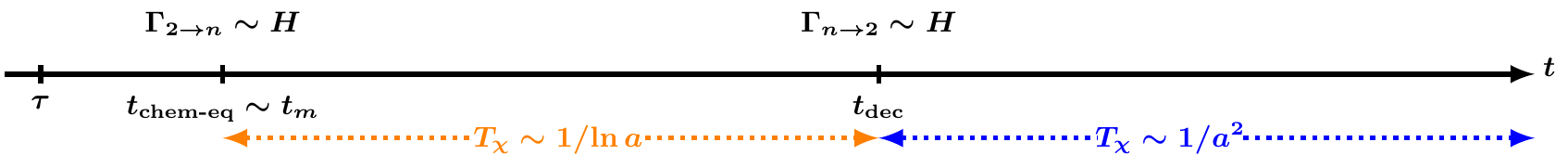} \label{fig:fig1b}}\\
\vspace{3mm}
\subfloat[Relativistic at $t_\text{chem-eq}$ and no cannibalism (RNC)]{\includegraphics[width = 0.8\textwidth]{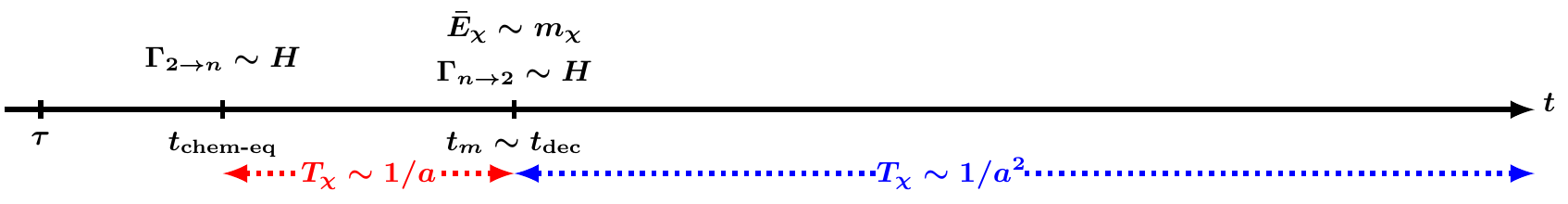}\label{fig:fig1c}}\\
\vspace{3mm}
\subfloat[Relativistic at $t_\text{chem-eq}$ and cannibalism (RC)]{\includegraphics[width = 0.8\textwidth]{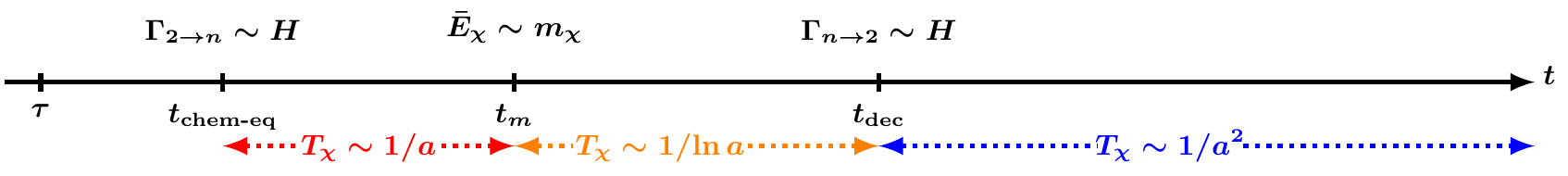}\label{fig:fig1d}} 
\caption{Four possible thermal histories of a self-interacting dark sector which is populated at the initial time $t=\tau$, reaches chemical equilibrium at $t=t_\text{chem-eq}$, becomes non-relativistic at $t=t_m$, and decouples and freezes out at $t=t_\text{dec}$. (a): The dark sector includes barely enough initial DM particles to reach chemical equilibrium; DM is non-relativistic at chemical equilibrium and then freezes out with no subsequent cannibal phase (NRNC). (b): The dark sector is initially populous enough to undergo a cannibal phase after establishing chemical equilibrium with non-relativistic particles and before decoupling (NRC).
(c): The dark sector is initially dense enough to reach chemical equilibrium while its particles are still relativistic, but not so dense as to experience a cannibal phase when it becomes non-relativistic (RNC). (d): The dark sector is initially so dense that it reaches chemical equilibrium when DM particles are still relativistic and it also undergoes a cannibal phase prior to freeze-out (RC). In each scenario, the temperature of the dark sector as a function of scale factor is also depicted; red shows the relativistic phase, orange corresponds to the cannibal phase and blue indicates the non-relativistic decoupled phase. Time intervals in the plot are for demonstrative purposes only and not indications of actual times.}
\label{fig:fig1}
\end{figure}

Fig.~\ref{fig:fig1} displays the different possibilities for the thermal history of a self-interacting dark sector after establishing chemical equilibrium. In Fig.~\ref{fig:fig1a}, we illustrate a thermal history for a scenario with the minimum initial number density of dark matter particles for which chemical equilibrium can be obtained;
here, the dark sector establishes chemical equilibrium when particles are non-relativistic and they decouple immediately without experiencing a cannibal phase. This scenario, which is non-relativistic at equilibrium with no cannibalism, is dubbed NRNC. Increasing the initial number density from this minimum value, eventually the dark sector becomes adequately populated such that chemical equilibrium is established by non-relativistic particles and followed by a cannibal phase before decoupling. This thermal history, non-relativistic at equilibrium followed by cannibalism (NRC), is displayed in Fig.~\ref{fig:fig1b}. An even larger initial number density of dark sector particles can lead to chemical equilibrium established by relativistic particles which will decouple without undergoing a cannibal phase (RNC), or even chemical equilibrium reached by relativistic particles, succeeded by a cannibal phase prior to decoupling (RC). These scenarios are pictured in Fig.~\ref{fig:fig1c} and Fig.~\ref{fig:fig1d} respectively.
  
Details of each of these four possible thermal histories, including characteristic moments and relevant temperatures, are covered in Appendix~\ref{sec:DSgeneral}.
 
\subsection{Thermal history of a self-interacting dark sector populated by Hawking evaporation of PBHs}

In the early Universe, during a radiation-dominated era, density fluctuations grow after they enter the cosmological horizon, and the overdense regions can collapse into PBHs. It is customary to represent
the initial abundance of PBHs by their energy density, $\rho_{\rm BH}$, normalized to the radiation (i.e.~visible sector) energy density, $\rho_{\rm V}$, at the time of their formation, $t_i$; this normalized initial PBH abundance is denoted by the dimensionless parameter $\beta \equiv \rho_{\rm BH}(t_i)/\rho_{\rm V}(t_i)$. Since the  energy
density of PBHs redshifts like the energy density of matter, an initially radiation-dominated universe will eventually become matter-dominated if the PBHs are long-lived enough. The critical initial abundance of PBHs corresponding to transitioning to an early matter (PBH)-dominated Universe is denoted by $\beta_\text{crit}(M_\text{BH})$. Hawking evaporation of PBHs completes at their lifetime, and their energy budget is emitted in the form of all particles in the spectrum which are lighter than the PBH temperature. 
The details of Hawking evaporation of PBHs, the spectra of emitted particles, and the condition for transitioning to an early PBH-dominated era have been reviewed in Appendix~\ref{sec:PBH}.

For simplicity, and to demonstrate the main conceptual features, we take the dark sector to consist of a self-interacting real scalar field, $\chi$, with Lagrangian
\begin{equation}
    \mathcal{L}=\frac{1}{2}\partial_\mu\chi\partial^\mu\chi-\frac{1}{2}m_\chi\chi^2-\frac{m_\chi\lambda}{3!}\chi^3 -\frac{\lambda^2}{4!}\chi^4,
    \label{eq:lagrangian}
\end{equation}
where $m_\chi$ is the mass of the DM particle and $\lambda$ represents the strength of self-interactions. In this study we are interested in models where interactions are strong, i.e.~$1\lesssim \lambda\lesssim4\pi$.

If $\beta<\beta_\text{crit}$, then at the time of evaporation, the Universe is radiation-dominated, and Hawking evaporation of PBHs populates the dark sector by emitting DM particles. The energy emitted by PBHs in the form of SM particles is negligible compared to the background radiation and equilibrates quickly with it. The emitted DM particles are relativistic, far from equilibrium, and their initial number density is proportional to the initial abundance of PBHs, $\beta$. 

Due to the sizeable self-interaction $\lambda$, the dark sector can evolve toward kinetic and chemical equilibrium.
It is reasonable to expect that for a fixed self-coupling, $\lambda$, there is an initial abundance of PBHs, denoted by $\beta_\text{kin}(m_\chi, M_\text{BH})$, such that for $0\lesssim\beta\lesssim\beta_\text{kin}$, the initial number density of DM particles is too small to establish kinetic equilibrium or, equivalently, a temperature for the dark sector. In this case, the rate of elastic scattering processes, $\chi\chi\leftrightarrow\chi\chi$, stays less than the Hubble expansion rate until DM particles become non-relativistic. Conversely, for any initial abundance of PBHs in the range, $\beta_\text{kin}\lesssim\beta\lesssim\beta_\text{crit}$, the dark sector will reach kinetic equilibrium before becoming non-relativistic.
After establishing kinetic equilibrium, number-changing processes, $\chi\chi\rightarrow\chi\chi\chi$, can drive the dark sector toward chemical equilibrium provided that their rate becomes comparable to the Hubble expansion rate while particles are still relativistic. 
As we show in Appendix~\ref{sec:DSgeneral}, establishing kinetic equilibrium by DM particles in a dark sector with a strong self-coupling guarantees that chemical equilibrium is established afterwards. This is a general statement that is also true for a dark sector populated by PBHs; therefore, $\beta\gtrsim\beta_\text{kin}$ results in chemical equilibrium.
 
Another threshold of the initial abundance of PBHs is denoted by $\beta_\text{NRC}(m_\chi,M_\text{BH})$, such that for $\beta_\text{kin}\lesssim\beta\lesssim\beta_\text{NRC}$, the emitted DM particles are abundant enough to establish chemical equilibrium, but  scarce enough that they reach chemical equilibrium when they are non-relativistic. They subsequently decouple without going through a cannibal phase (we refer to this case as non-relativistic then no-cannibalism, or NRNC).

The next reasonable threshold is indicated by $\beta_\text{rel}(m_\chi,M_\text{BH})$, such that for $\beta_\text{NRC}\lesssim\beta\lesssim\beta_\text{rel}$, the number density of emitted DM particles is small enough that the dark sector reaches chemical equilibrium when particles are non-relativistic, but also large enough that the dark sector undergoes a subsequent cannibal phase before decoupling (we refer to this case as non-relativistic then cannibalism, or NRC).

By increasing the initial abundance of the dark sector, the next important threshold of the initial abundance of PBHs, denoted by $\beta_\text{RC}(m_\chi,M_\text{BH})$, is reached such that for $\beta_\text{rel}\lesssim\beta\lesssim\beta_\text{RC}$, the dark sector reaches chemical equilibrium while DM particles are still relativistic; however, it does not experience a subsequent cannibal phase when DM particles become non-relativistic (we refer to this case as relativistic then no-cannibalism, or RNC).

Finally for $\beta_\text{RC}\lesssim\beta\lesssim\beta_\text{crit}$ the dark sector reaches chemical equilibrium when its particles are still relativistic and it certainly undergoes a cannibal phase when DM particles become non-relativistic (we refer to this case as relativistic then cannibalism, or RC). Needless to say, the thresholds for the initial abundance of PBHs mentioned above consistently satisfy the inequality
\begin{equation} 
    \beta_\text{kin}\lesssim\beta_\text{NRC}\lesssim\beta_\text{rel}\lesssim\beta_\text{RC}\lesssim\beta_\text{crit},
    \label{betathresh}
\end{equation}
where $\beta_\text{kin}$, $\beta_\text{NRC}$, $\beta_\text{rel}$, $\beta_\text{RC}$, and $\beta_\text{crit}$ are provided analytically by Eqs.~(\ref{eq:betakin}), (\ref{eq:betacannnonrel}), (\ref{eq:betarel}),  (\ref{eq:betacannrel}), and (\ref{eq:betac}), respectively. 
 
 It is also possible that the initial abundance of PBHs happens to be larger than the critical value, $\beta\gtrsim\beta_\text{crit}$. In that case, the PBHs initiate an early matter (PBHs)-dominated era before their evaporation, and they reheat both the visible sector and the dark sector with their Hawking radiation. The abundance of emitted particles is independent of $\beta$~\cite{Baumann:2007yr}, and different thermal histories for the dark sector are defined by different thresholds for the PBH mass, each dependent on the DM mass, $m_\chi$.

Since $\beta_\text{crit}\sim M_\text{Pl}/M_\text{BH}$, there exists a PBH mass threshold $M_\text{BH,kin}$ such that for $M_\text{BH}\lesssim M_\text{BH,kin}$ kinetic equilibrium and chemical equilibrium in the dark sector are assured, while for $M_\text{BH}\gtrsim M_\text{BH,kin}$ the emitted DM particles are not abundant enough to reach kinetic equilibrium. The next expected threshold is $M_\text{BH,NRC}$, for which PBHs with a mass in the $M_\text{BH,NRC}\lesssim M_\text{BH}\lesssim M_\text{BH,kin}$ range lead to an NRNC thermal history\footnote{We note that we have chosen to keep the threshold subscripts as in the radiation-dominated case, though, due to the inverse relationship between $\beta_{\rm crit}$ and $M_{\rm BH}$, the thermal histories are flipped for the matter-dominated case.}. 
It is reasonable to assume that another threshold can be defined as $M_\text{BH,rel}$ where $M_\text{BH,rel}\lesssim M_\text{BH}\lesssim M_\text{BH,NRC}$ gives rise to an NRC thermal history. The next  possible threshold is indicated as $M_\text{BH,RC}$ where $M_\text{BH,RC}\lesssim M_\text{BH}\lesssim M_\text{BH,rel}$ is responsible for an RNC thermal history; finally, a PBH mass in the range $M_\text{BH}\lesssim M_\text{BH,RC}$ initiates an RC thermal history.

The following inequality holds for these thresholds consistently:
\begin{equation}
  M_\text{BH,RC}\lesssim M_\text{BH,rel} \lesssim M_\text{BH,NRC} \lesssim M_\text{BH,kin},
  \label{MBHthresh}
\end{equation}
where $M_\text{BH,rel}$, $M_\text{BH,NRC}$, and $M_\text{BH,kin}$ are given analytically by Eqs.~(\ref{eq:MBHrel}), (\ref{eq:MBHcannnonrel}), and (\ref{eq:MBHkin}), respectively.

Table~\ref{tab:scenarios} summarises all possible thermal histories of a dark sector populated by Hawking evaporation of PBHs in an early radiation-dominated (second column) and an early matter-dominated (third column) Universe.

\begin{table*}
\centering
\begin{tabular}{|c|c|c|}
\hline
Thermal History & Early Radiation Domination & Early Matter Domination\\
\hline \hline
NRNC (non-relativistic, no cannibalism)  &$\beta_\text{kin}\lesssim\beta\lesssim\beta_\text{NRC}$&$M_\text{BH,NRC}\lesssim M_\text{BH}\lesssim M_\text{BH,kin}$\\
NRC (non-relativistic, cannibalism)&$\beta_\text{NRC}\lesssim\beta\lesssim\beta_\text{rel}$&$M_\text{BH,rel}\lesssim M_\text{BH}\lesssim M_\text{BH,NRC}$\\
RNC (relativistic, no cannibalism) & $\beta_\text{rel}\lesssim\beta\lesssim\beta_\text{RC}$&$M_\text{BH,RC}\lesssim M_\text{BH}\lesssim M_\text{BH,rel}$ \\
RC (relativistic, cannibalism)& $\beta_\text{RC}\lesssim\beta\lesssim\beta_\text{crit}$&$M_\text{BH}\lesssim M_\text{BH,RC}$ \\
\hline 
\end{tabular}
\caption{Different possible thermal histories of a self-interacting dark sector populated by Hawking evaporation of PBHs. The second column which corresponds to the case where PBHs evaporates within a radiation-dominated era, shows the relevant range of initial abundance of PBHs that can lead to a specific thermal history. The third column contains different possible ranges of the mass of PBHs that lead to different thermal histories when PBHs dominate the energy density of the Universe prior to their evaporation.}
\label{tab:scenarios}
\end{table*}

\section{Results}
\label{sec:results}

In this section, we present our results for the possible thermal histories resulting from the population of a strongly self-interacting dark sector by Hawking evaporation of PBHs.
The relevant model parameters are the initial abundance of PBHs, $\beta$, the PBH mass, $M_\text{BH}$, and the DM mass, $m_\chi$. We assume that the self-coupling $\lambda\simeq 1$ throughout; the main conclusions are unchanged as long as the self-coupling is within the strong interaction range.  

\begin{figure}[t]
  \centering
  \includegraphics[width=0.45\textwidth]{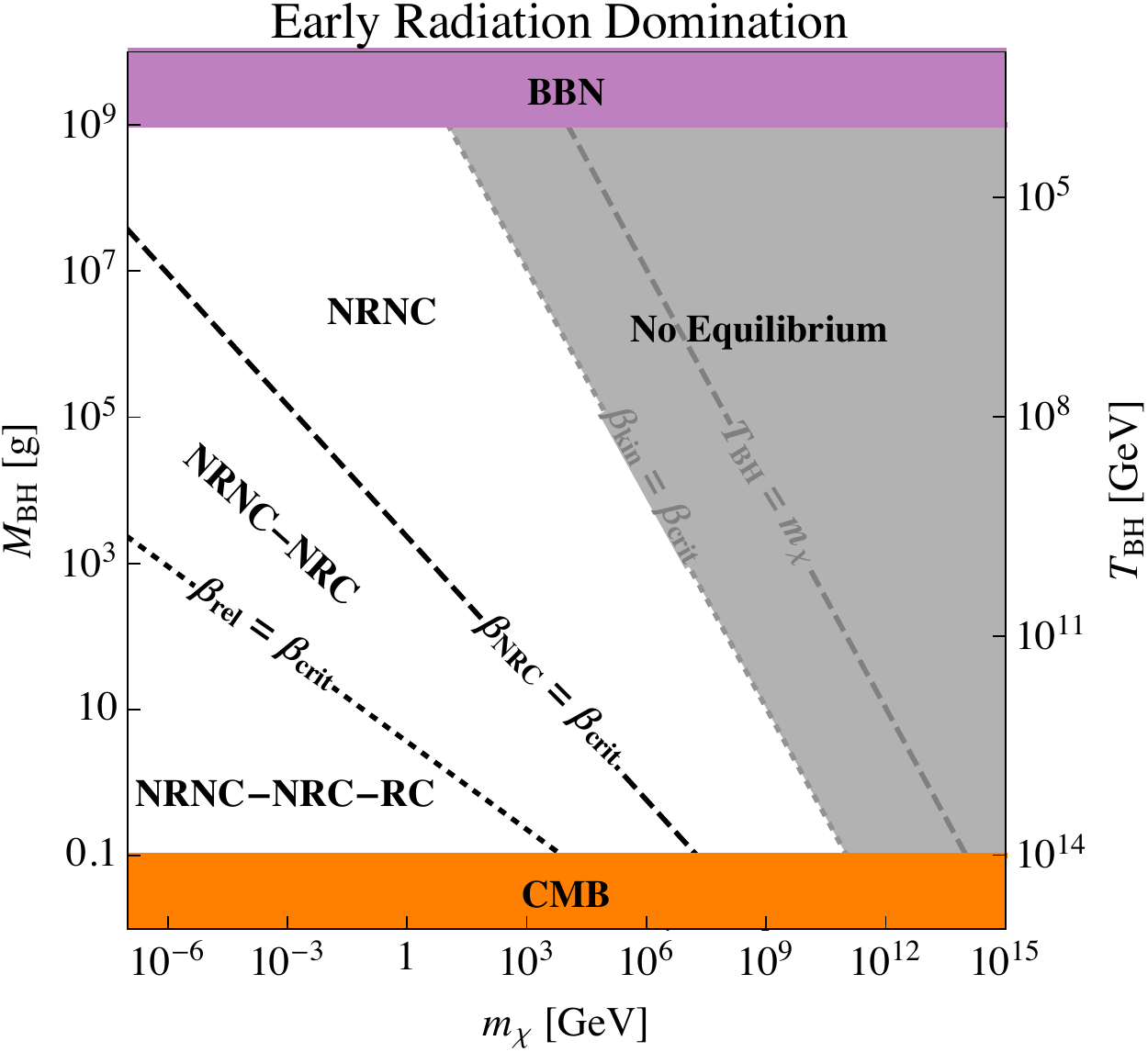}
  \includegraphics[width=0.45\textwidth]{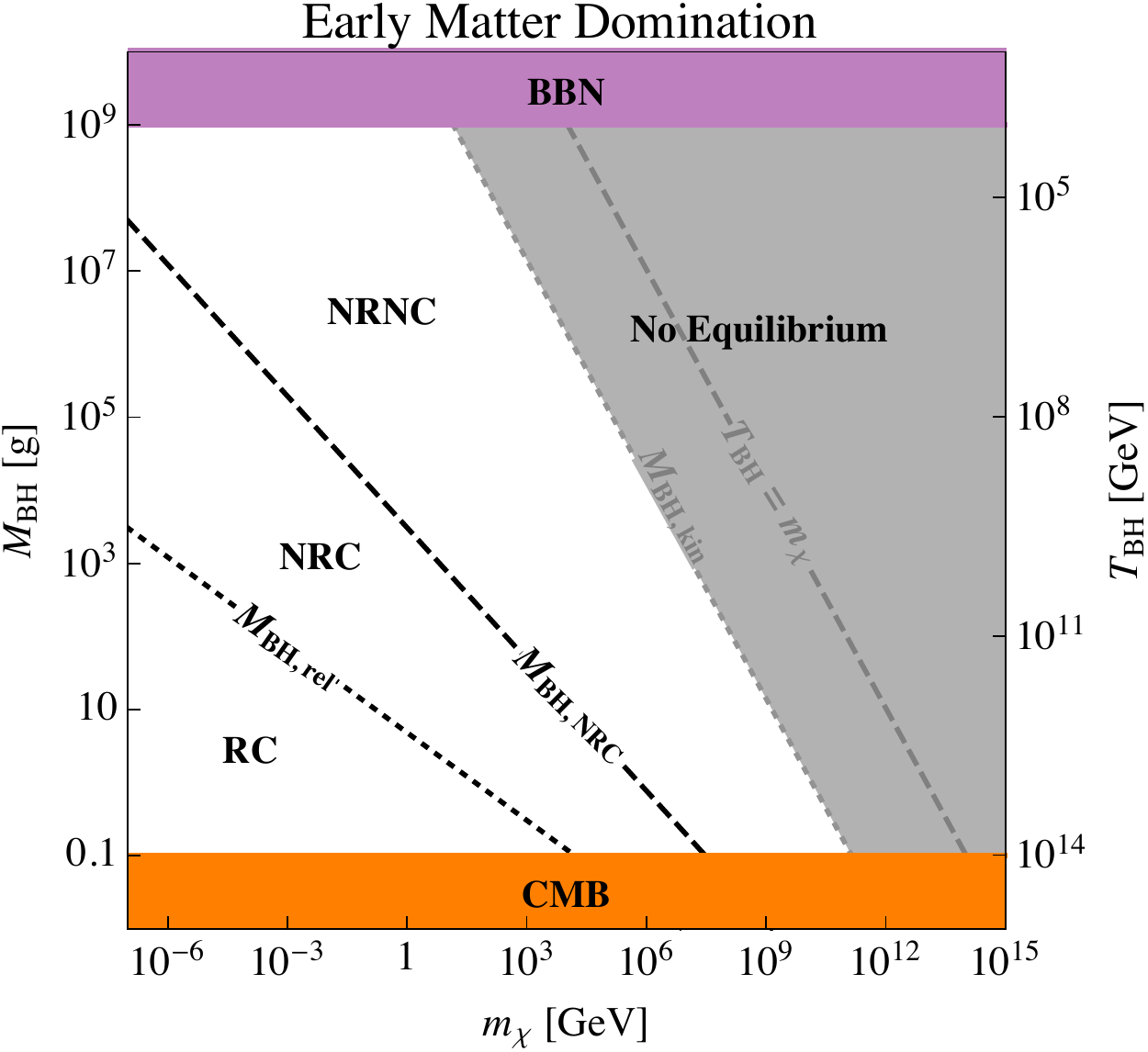}
  \caption{Different possible thermal histories of a self-interacting dark sector populated by Hawking evaporation of  PBHs in an early radiation-dominated Universe, $\beta<\beta_\text{crit}$ (left) and a matter-dominated Universe, $\beta\geq\beta_\text{crit}$ (right) in the $(m_\chi, M_\text{BH})$ plane, with regions demarcated by the various abundance thresholds in Eqs.~(\ref{betathresh}) and (\ref{MBHthresh}). In both panels, when the initial temperature of PBHs is smaller than the mass of DM particles (above the grey dashed line marked by $T_\text{BH}=m_\chi$), the emitted DM particles are not abundant enough to thermalize before becoming non-relativistic. When the initial temperature of PBHs is larger than the DM mass (below the grey dashed line, $T_\text{BH}=m_\chi$), thermalization can still be unreachable. This region is bounded by the dotted grey line tagged by $\beta_\text{kin}=\beta_\text{crit}$ ($M_\text{BH,kin}$) in the left (right) panels. 
  In the left panel, the region below the dotted grey line labeled by $\beta_\text{kin}=\beta_\text{crit}$ can accommodate an NRNC thermal history. Below the black dashed line, marked by $\beta_\text{NRC}=\beta_\text{crit}$, NRNC and NRC can both occur. The RC thermal history can only take place below the black dotted line labeled by $\beta_\text{rel}=\beta_\text{crit}$. In the right panel, 
  below the black dotted line labeled $M_\text{BH,rel}$, the only possible thermal history is RC. Between the black dotted line and the black dashed line labeled $M_\text{BH,NRC}$, the only feasible thermal history is NRC. Above the black dashed line and below the grey dotted line labeled $M_\text{BH,kin}$, the only possible thermal history is NRNC. In both panels, the RNC thermal history can never be realized.}
  \label{fig:EQregions}
\end{figure}

\begin{figure}[ht]
  \centering
  \includegraphics[width=0.45\textwidth]{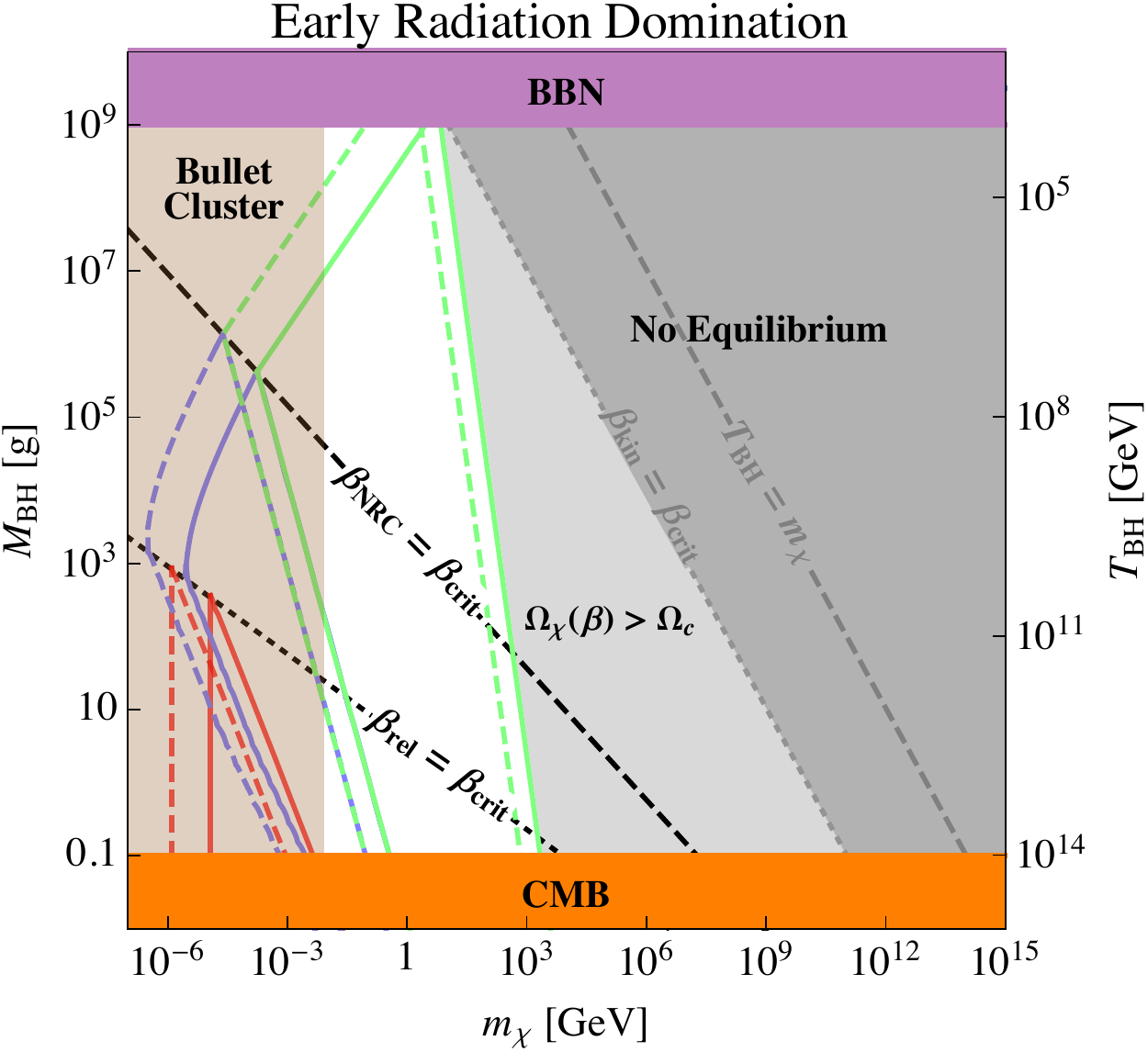}
  \includegraphics[width=0.45\textwidth]{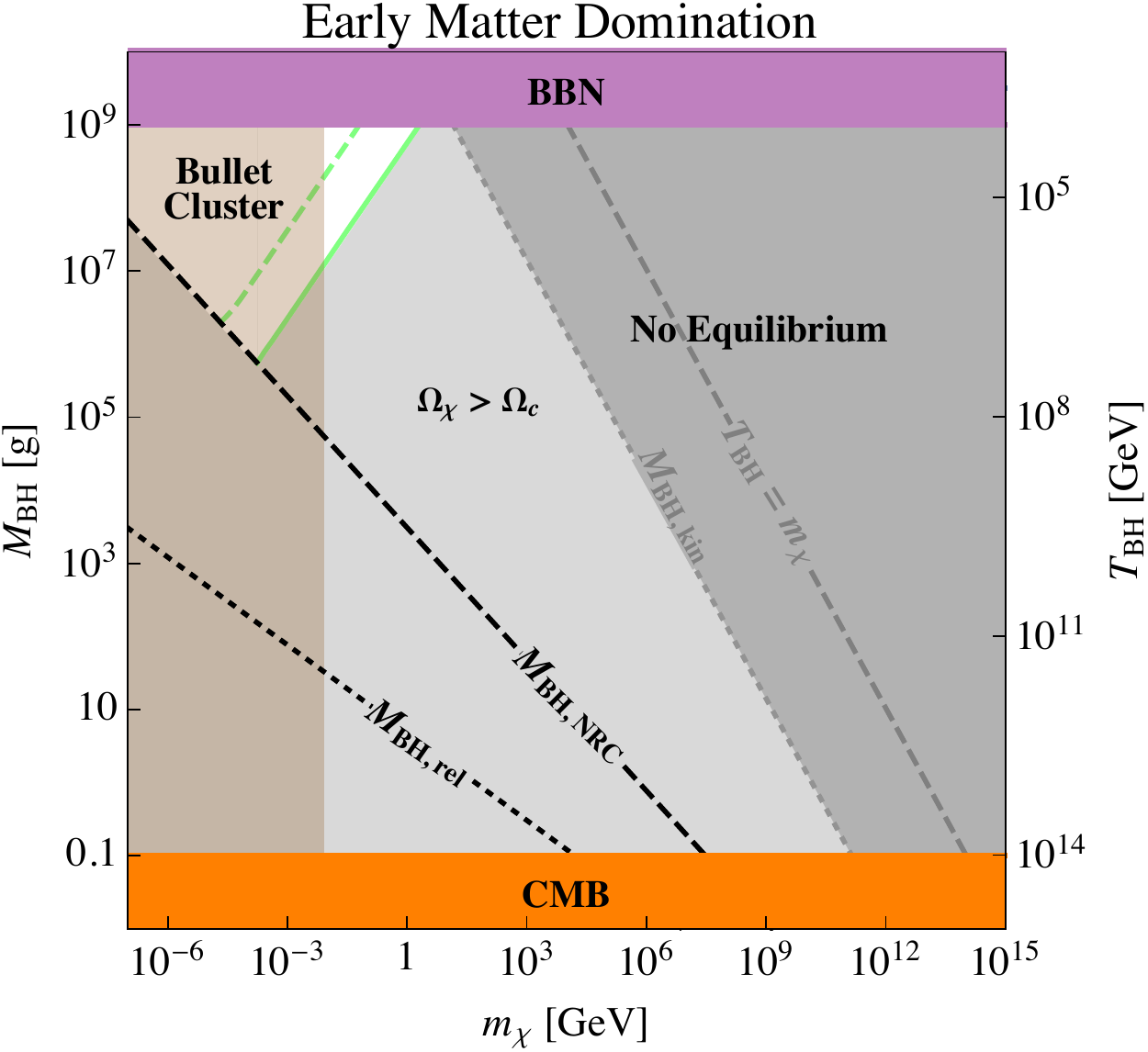}
  \caption{Constraints on equillibrated self-interacting dark sector populated by Hawking evaporation of  PBHs in an early radiation-dominated Universe, $\beta<\beta_\text{crit}$ (left) and a matter-dominated Universe, $\beta\geq\beta_\text{crit}$ (right) in the $(m_\chi, M_\text{BH})$ plane. The brown region depicts the bound on DM self–interaction from the Bullet Cluster. The light grey area corresponds to region where all viable thermal histories (NRNC, NRC, and RC) overproduce DM ($\Omega_\chi>\Omega_c$). In the left panel, inside the region enclosed by green solid boundary, the NRNC thermal history, within the region enclosed by blue solid boundary, the NRC thermal history, and inside the region enclosed by red solid boundary, the RC thermal history, leads to the right DM relic abundance today.  inside the regions enclosed by dashed boundaries in the left panel, the three viable thermal histories lead to underproduction of DM by a factor of 10. In the right panel, the NRC and RC thermal histories overproduce DM. The green solid (dashed) line specifies PBH and DM masses that result in the right DM relic abundance ( underproduction of DM by a factor of 10) today via NRNC scenario.}
  \label{fig:abundance}
\end{figure}

In Figs.~\ref{fig:EQregions} and \ref{fig:abundance} we present the possible thermal histories of the dark sector in the $(m_\chi, M_\text{BH})$ plane, with regions demarcated by the various abundance thresholds in Eqs.~(\ref{betathresh}) and (\ref{MBHthresh}). 
The left and right panels of Figs.~\ref{fig:EQregions} and \ref{fig:abundance} correspond to an early radiation-dominated ($\beta<\beta_\text{crit}$) and matter-dominated ($\beta\geq\beta_\text{crit}$) Universe, respectively.  In each panel, the lower and  upper bounds on the PBH mass, $0.1\,\text{g}\lesssim M_\text{BH}\lesssim 10^9\,\text{g}$, are imposed by CMB and BBN constraints, respectively~\cite{Gondolo:2020uqv}. As we show in Appendix~\ref{sec:DS}, in both panels of Fig.~\ref{fig:EQregions}, when the initial temperature of PBHs is smaller than the mass of DM particles (above the grey dashed line marked by $T_\text{BH}=m_\chi$), the emitted DM particles are not abundant enough to thermalize before becoming non-relativistic. This feature persists even when the  self-coupling, $\lambda$, saturates the perturbative unitarity limit. When the initial temperature of PBHs is larger than the DM mass (below the grey dashed line, $T_\text{BH}=m_\chi$), thermalization can still be unattainable. This region is bounded by the dotted grey line  tagged by $\beta_\text{kin}=\beta_\text{crit}$ ($M_\text{BH,kin}$) in the left (right) panels. Above this line, (grey region), in left (right) panel, thermalization (and therefore chemical equilibrium) is not feasible. 

As is evident from the left panel of Fig.~\ref{fig:EQregions}, the region below the dotted grey line labeled by $\beta_\text{kin}=\beta_\text{crit}$ can accommodate an NRNC thermal history. Below the black dashed line, marked by $\beta_\text{NRC}=\beta_\text{crit}$, NRNC and NRC can both occur. In contrast, the RC thermal history can only take place below the black dotted line labeled by $\beta_\text{rel}=\beta_\text{crit}$; in this regime, all three histories (NRNC, NRC, and RC) can occur. We find that the RNC history can never be realized; for a dark sector which is populated by Hawking evaporation of PBHs in an early radiation-dominated Universe and which reaches chemical equilibrium when its particles are still relativistic, a cannibal phase is inevitable. 

In the right panel of Fig.~\ref{fig:EQregions}, we show the corresponding results when PBHs dominate the energy density of the Universe. Below the black dotted line labeled $M_\text{BH,rel}$, the only possible thermal history is RC. Between the black dotted line and the black dashed line labeled $M_\text{BH,NRC}$, the only feasible thermal history is NRC.  And finally, above the black dashed line and below the grey dotted line labeled $M_\text{BH,kin}$, the only possible thermal history is NRNC. 

In Fig.~\ref{fig:abundance}, we again show the possible thermal histories on the same plane as Fig.~\ref{fig:EQregions}, but now with additional information about the DM relic density for the different cases, as well as the constraints coming from the Bullet Cluster. As before, the left and right panels correspond to an early radiation-dominated ($\beta<\beta_\text{crit}$) and matter-dominated ($\beta\geq\beta_\text{crit}$) Universe, respectively. The DM relic abundance today is denoted by $\Omega_\chi$, with $\Omega_\chi=\Omega_c$ being the value observed today.  In both panels of Fig.~\ref{fig:abundance}, the light grey area corresponds to region where $\Omega_\chi>\Omega_c$, and therefore whatever the thermal history of the equilibrated dark sector, this region of parameter space overcloses the Universe. 
The bound on DM self–interactions from the Bullet Cluster, $\sigma_{\chi\chi}/m_\chi\lesssim 1\,\text{cm}^2/\text{g}$~\cite{Randall:2008ppe}, where $\sigma_{\chi\chi}=\lambda^4/(128\pi m_\chi^2)$ is the DM self-scattering cross section, leads to a lower bound on the DM mass for a fixed self-coupling, and is displayed as the excluded brown region ($m_\chi\lesssim 8\, \text{MeV}$) in Fig.~\ref{fig:abundance}.

In the left panel of Fig.~\ref{fig:abundance}, we show regions in which each of the three viable thermal histories, NRNC, NRC, and RC, gives rise to the observed DM relic abundance today.
The parameter space for which $\Omega_\chi(\beta)=\Omega_c$ (i.e.~there exists an allowed value of $\beta$ such that this is possible) is outlined in a solid contour for each thermal history: NRNC (green), NRC (blue), and RC (red).
The corresponding dashed lines of each color enclose regions with $\Omega_\chi(\beta)=0.1 \times \Omega_c$ for each thermal history. The NRC thermal history yields the right DM relic abundance today for DM with a mass in the $8\,\text{MeV}\lesssim m_\chi\lesssim 360\,\text{MeV}$ range if the PBH mass is in the $0.1\,\text{g}\lesssim M_\text{BH}\lesssim 205\,\text{g}$ range, while the NRNC thermal history leads to the observed DM relic abundance today for DM with a mass in the $8\,\text{MeV}\lesssim m_\chi\lesssim 2.2\,\text{TeV}$ range for PBH masses in the range $0.1\,\text{g}\lesssim M_\text{BH}\lesssim 10^{9}\,\text{g}$. The reason that the NRC thermal history prefers light DM while the NRNC thermal history permits heavier DM can be understood by considering that the common boundary of the NRC and NRNC regions corresponds to $\beta_\text{NRC}$ leading to the right relic abundance today, and it is an increasing function of both $m_\chi$ and $M_\text{BH}$ (see Eq.~(\ref{eq:betacannnonrel})).
One can see that the NRC and RC regions overlap, which means that within the overlapping region, there are two different  thermal histories (and equivalently two different $\beta$s) that yield the observed relic abundance today. It is worth mentioning, however, that the Bullet Cluster excludes the RC scenario altogether. 

The right panel of Fig.~\ref{fig:abundance} follows the same color conventions as the left panel. One sees that for an early matter (PBH)-dominated Universe, the only thermal history that does not overproduce DM is NRNC. 

In Appendix~\ref{sec:DS}, we present the full calculation of relevant temperatures of the dark and visible sectors.  In particular, we calculate the temperature of the dark sector at the time of chemical equilibrium and compare it with the temperature of the visible sector at the same time for some benchmark DM masses. The ratio $\xi$ of temperatures at chemical equilibrium is computed after imposing the constraint of the correct relic density in the current Universe. We find that when PBHs evaporate in a radiation-dominated Universe, NRNC and NRC thermal histories can lead to dark sectors that are colder ($\xi<1$) or hotter ($\xi>1$) than the visible sector. For an RC thermal history, however, only $\xi<1$ is possible. When PBHs dominate the energy density of the Universe before their evaporation, NRNC and NRC thermal histories always lead to a hotter dark sector ($\xi>1$); an RC thermal history, on the other hand, always leads to $\xi=1$.

\begin{figure}[ht]
  \centering
  \includegraphics[width=0.4\textwidth]{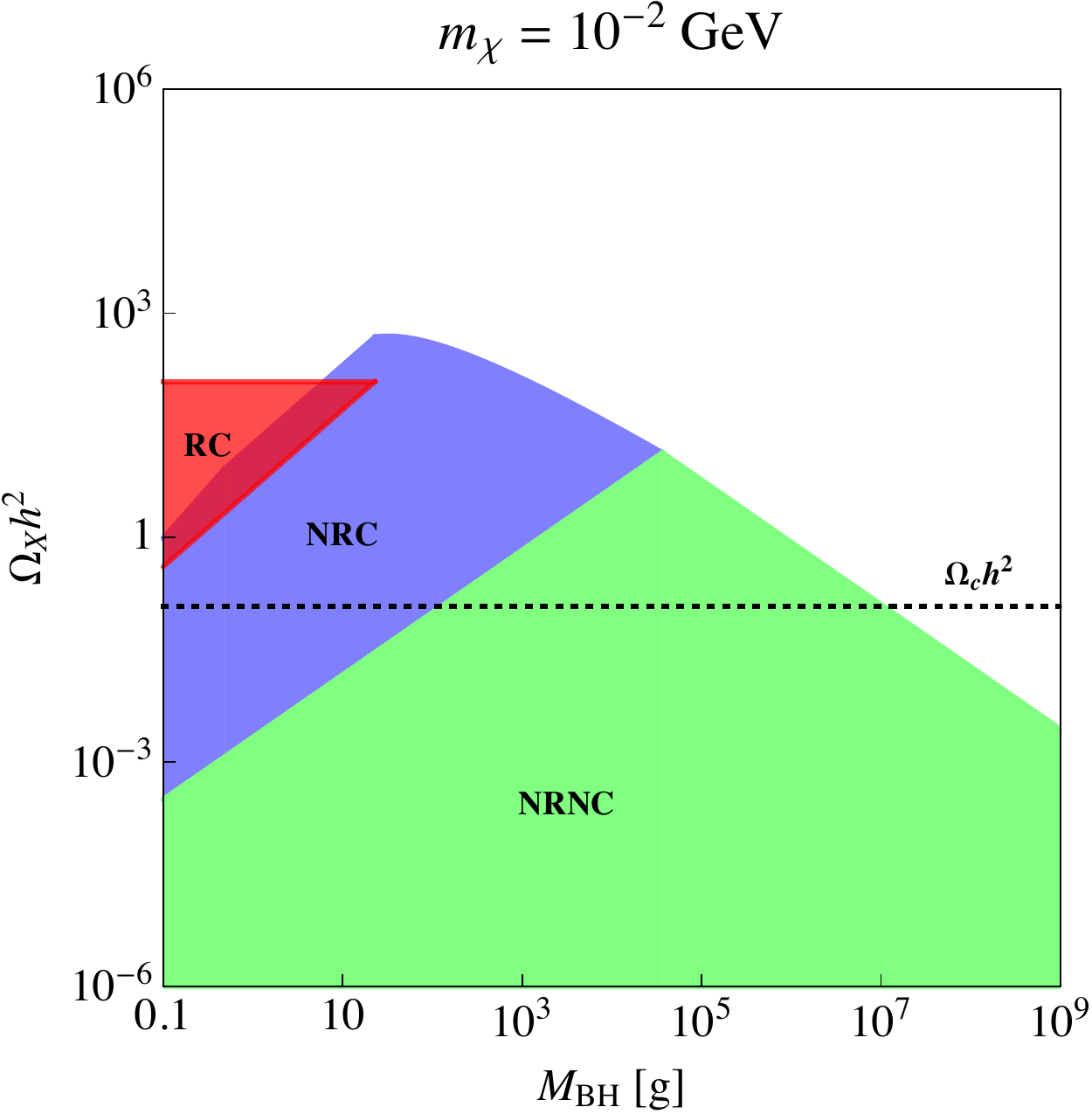}
  \includegraphics[width=0.445\textwidth]{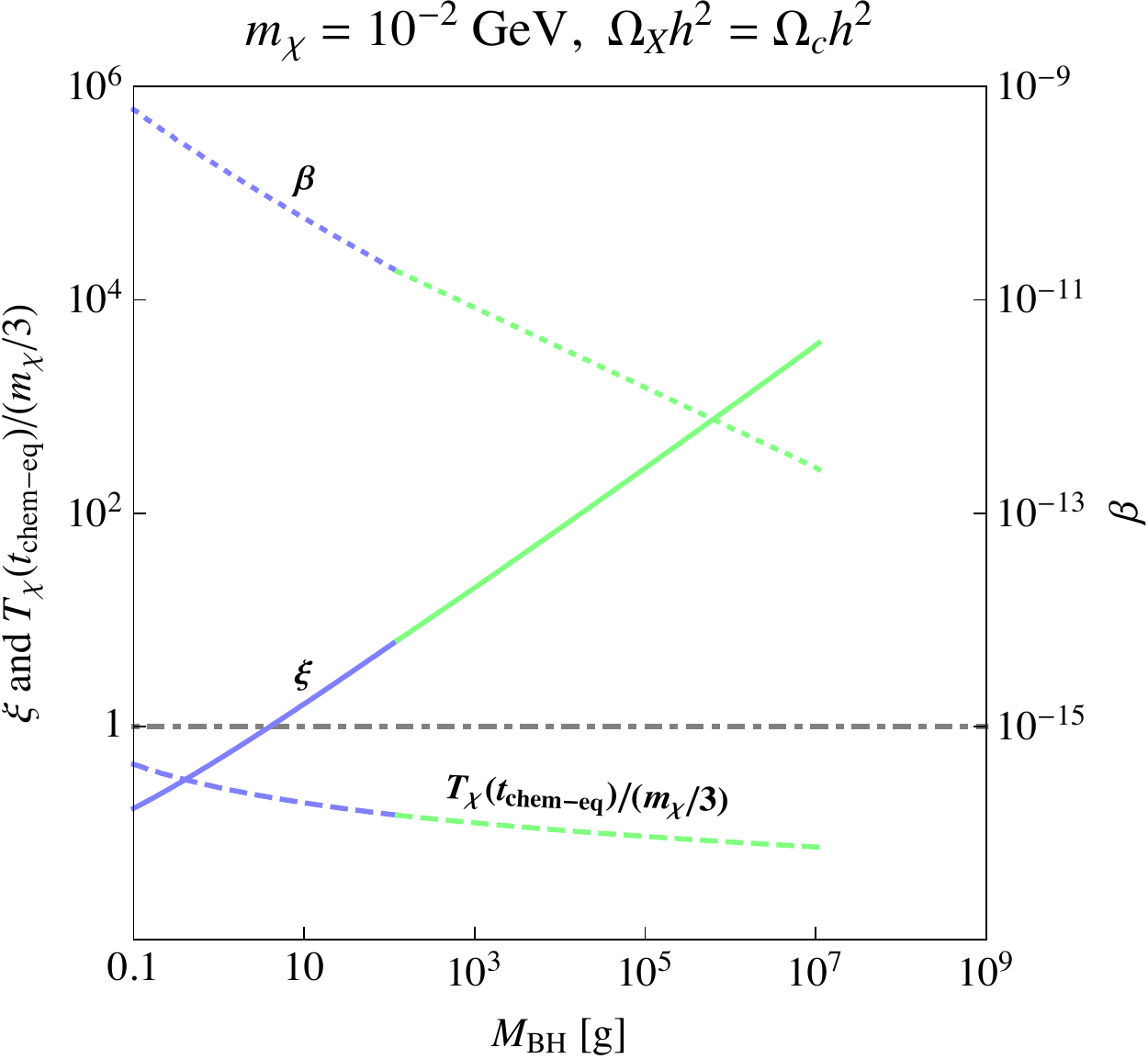}\\
  \includegraphics[width=0.4\textwidth]{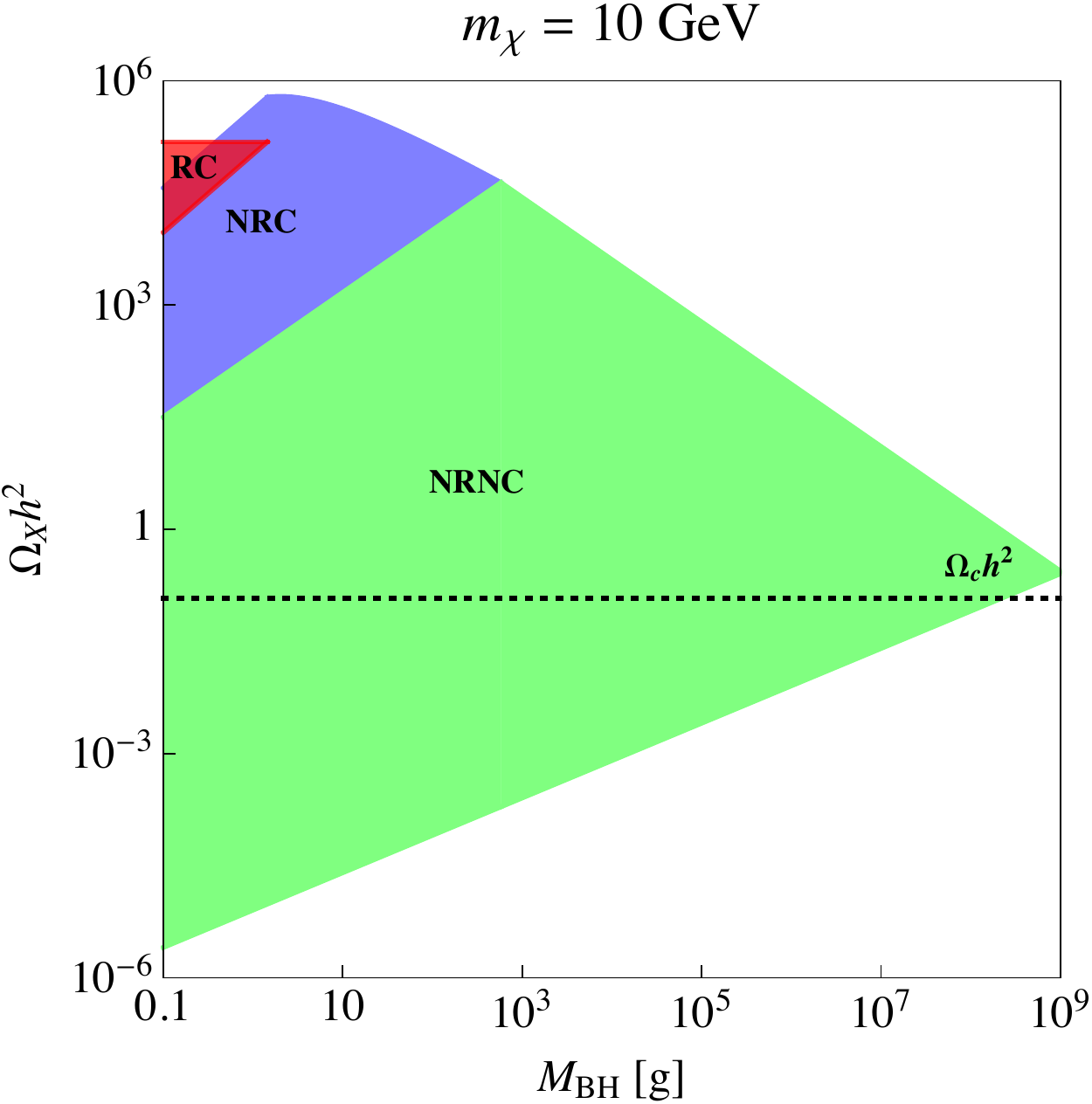}
  \includegraphics[width=0.445\textwidth]{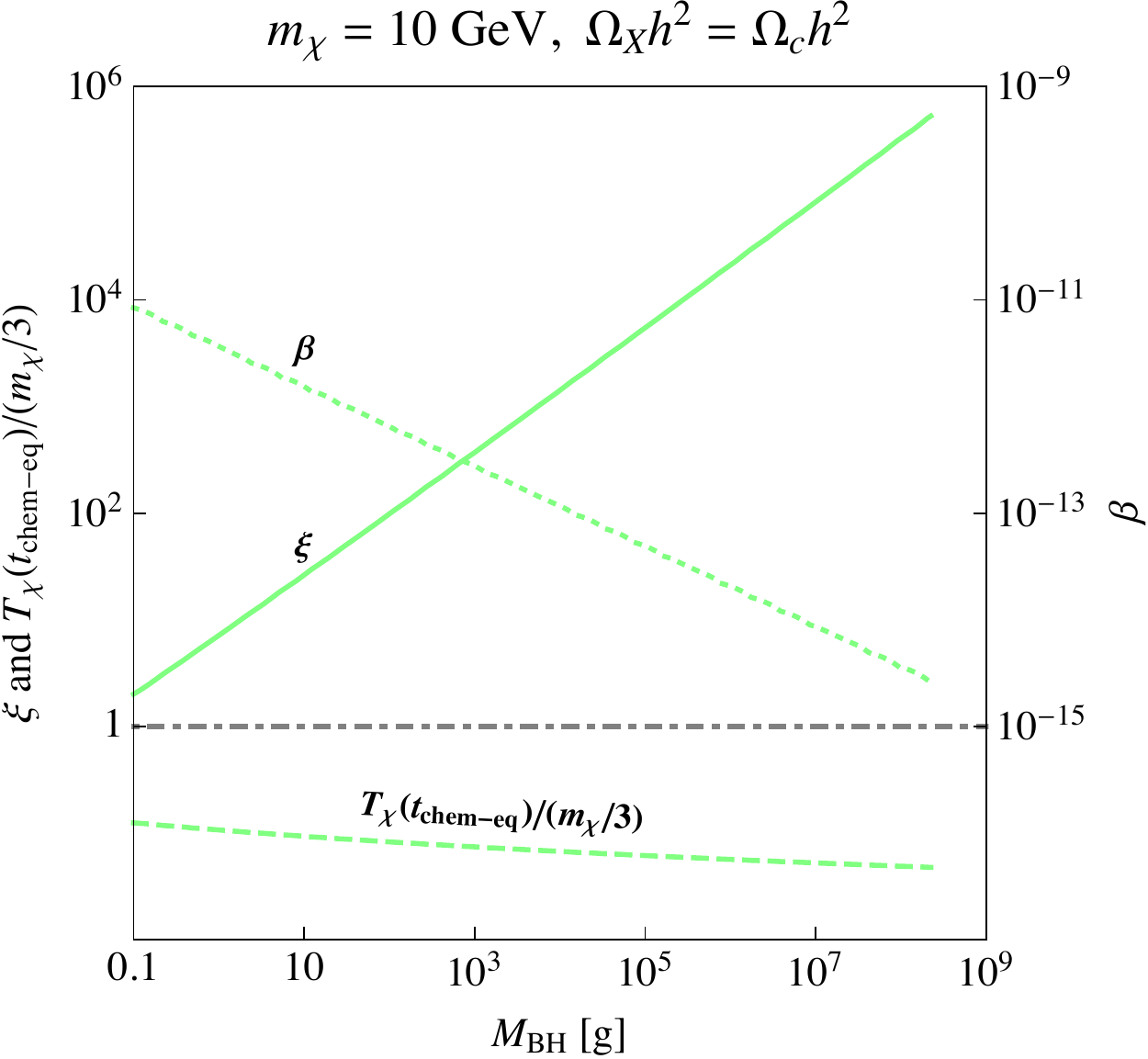}
  \caption{Top (Bottom) panel corresponds to the DM mass of $m_\chi=10^{-2}\,\text{GeV}$ ($m_\chi=10\,\text{GeV}$). Left panels show the relic abundance today originated from an RC thermal history (red), an NRC thermal history (blue), and and NRNC thermal history (green). Points in white regions in left panels cannot be reached by any of these three viable scenarios. Right panels depict ratio of the temperature of the dark sector at chemical equilibrium to the temperature of the visible sector at that time, $\xi$ (solid), the ratio of the temperature of the dark sector to one third of the mass of DM, $T_\chi(t_\text{chem-eq})/(m_\chi/3)$ (dashed) as an indicator of DM particles being relativistic ($T_\chi(t_\text{chem-eq})/(m_\chi/3)\gtrsim1$) or non-relativistic ($T_\chi(t_\text{chem-eq})/(m_\chi/3)<1$) at equilibrium time, and the initial abundance of PBHs, $\beta$, (dotted), assuming the dark sector has the right relic abundance today.}
  \label{fig:temps}
\end{figure}

In Fig.~\ref{fig:temps}, we show the DM relic abundance today (left panels) as well as relevant temperature ratios and the PBH abundance associated with the observed DM abundance (right panels) for two benchmark DM masses, $m_\chi=10^{-2}\,\text{GeV}$ (top panels) and $m_\chi=10\,\text{GeV}$ (bottom panels). The relic density is displayed as a function of the PBH mass on the left panels of Fig.~\ref{fig:temps} for the radiation-dominated case. The color conventions are the same as those in Fig.~\ref{fig:abundance}, with green, blue, and red corresponding to thermal histories NRNC, NRC, and RC, respectively. Regions that are white cannot be fulfilled by any thermal history.

For each benchmark DM mass, the right panel of Fig.~\ref{fig:temps} displays several quantities, all evaluated such that $\Omega_\chi(\beta)= \Omega_c$. The solid lines correspond to the ratio of temperatures $\xi$. The dashed lines correspond to the ratio of the temperature of the dark sector at chemical equilibrium to one third of the mass of DM, $T_\chi(t_\text{chem-eq})/(m_\chi/3)$. This is a diagnostic of DM being relativistic ($T_\chi(t_\text{chem-eq})/(m_\chi/3)\gtrsim1$) or non-relativistic ($T_\chi(t_\text{chem-eq})/(m_\chi/3)<1$) at chemical equilibrium. The dotted lines correspond to the initial PBH abundance, $\beta$. Here, the color conventions are as in the left panels: NRNC (green), and NRC (blue). 
For $m_\chi=10^{-2}\,\text{GeV}$, PBHs with a mass in the range $0.1\,\text{g}\lesssim M_\text{BH}\lesssim 1.2 \times 10^2\,\text{g}$ yield the right relic abundance of DM today through NRC thermal history, while PBH masses in the $1.2 \times 10^2\,\text{g}\lesssim M_\text{BH}\lesssim 10^7\,\text{g}$ range yield the right DM abundance today through the NRNC thermal history. The crossing point of these two cases ($M_\text{BH}= 1.2\times 10^2\,\text{g}$) corresponds to $\beta=\beta_\text{NRC}$.
For $m_\chi=10\,\text{GeV}$, PBHs in the mass range $0.1\,\text{g}\lesssim M_\text{BH}\lesssim 2.2 \times 10^8\,\text{g}$ yield the observed DM abundance via the NRNC thermal history.

While figures equivalent to the right panels of Fig.~\ref{fig:temps} are not displayed  for the matter-dominated case, the results for that case are evident from the right panel of Fig.~\ref{fig:abundance}.
The NRNC thermal history for DM with a mass in the $8\,\text{MeV}\lesssim m_\chi\lesssim 2.1\,\text{GeV}$ range leads to the right relic abundance today which is also consistent with the Bullet Cluster. The corresponding end points, $m_\chi=8\,\text{MeV}$ and $M_\text{BH}=1.2\times 10^7\,\text{g}$, and $m_\chi=2.1\,\text{Gev}$ and $M_\text{BH}=10^9\,\text{g}$, lead to  $\xi\simeq3.3\times 10^3$ and $\xi\simeq5.4 \times10^5$, respectively, and the ratio of the temperature of the dark sector to one third of the DM mass, $T_\chi(t_\text{chem-eq})/(m_\chi/3)$, is $0.08$ and $0.05$, respectively.

We also note that a temperature asymmetry larger than one, or equivalently a dark sector hotter than the visible sector, should not be considered problematic since it always happens when DM particles are non-relativistic and, accordingly, the dark sector should be considered a cold sector.  Furthermore, the contribution of the dark sector to the energy density of the Universe in this case is still negligible compared to that of the visible sector. We should add that the parameter space for this minimal dark sector is not constrained by $\Delta N_\text{eff}$, as is expected from previous studies of constraints on scalar particles emitted by non-spinning PBHs by $\Delta N_\text{eff}$~\cite{Hooper:2019gtx, Masina:2020xhk}.

\section{CONCLUSIONS}
\label{sec:conclusion}
In this paper we have explored the Hawking evaporation of light PBHs as a novel mechanism to reheat dark sectors with purely gravitational coupling to the visible sector. Operators involving the inflaton and visible and dark sector fields, which could potentially thermalize the two sectors in usual reheating scenarios, are not relevant to our scenario. In the absence of non-gravitational mediators between two sectors, any temperature asymmetry between them is persistent and evolves to keep the entropy of each sector conserved during the expansion of the Universe.

Taking a simple dark sector with a single self-interacting real scalar field, we have explored the possible thermal histories of the dark sector. The histories depend on the initial abundance of PBHs, the PBH mass, and the mass of the DM, and have been displayed in Figures \ref{fig:EQregions}, \ref{fig:abundance}, and \ref{fig:temps}. We have shown that when the DM mass is larger than the initial temperature of PBHs, the dark sector cannot establish kinetic and consequently chemical equilibrium. Of the four possible thermal histories, we have found that consistency conditions rule out the RNC history.

We find that for evaporation in a radiation-dominated Universe, the bound from the Bullet Cluster completely excludes the RC thermal history. The NRC and NRNC thermal histories can yield the right relic abundance today for DM with mass in the range $8\,\text{MeV}\lesssim m_\chi\lesssim 2.2\,\text{TeV}$, for PBH masses in the range  $0.1\,\text{g}\lesssim M_\text{BH}\lesssim 10^9\,\text{g}$. For evaporation in a matter (PBH)-dominated Universe, the RC and NRC thermal histories always overproduce DM, and only the NRNC thermal history can produce the right relic abundance today. This mechanism is viable for DM with a mass in the range $8\,\text{MeV}\lesssim m_\chi\lesssim 2.1\,\text{GeV}$, for PBH masses in the $1.2\times 10^7\,\text{g}\lesssim M_\text{BH}\lesssim 10^9\,\text{g}$ range.

We have also studied the ratio of the temperature of the dark to that of the visible sector, $\xi$. For  an early radiation-dominated Universe, the NRNC and NRC thermal histories can accommodate  $\xi$ less than, equal to, or greater than one. For RC thermal histories, on the other hand, only $\xi < 1$ is allowed. For an early matter (PBH)-dominated Universe, an NRNC or NRC thermal history forces $\xi > 1$, while an RC thermal history forces $\xi = 1$.

It is well known that light PBHs, motivated by numerous scenarios, are potential sources of DM. Here, we have demonstrated that a dark sector that is only gravitationally coupled to the visible sector and is populated by Hawking evaporation of PBHs may reach chemical equilibrium via sizeable self-interaction. Equilibration and the DM relic abundance today are restrictive enough to make this reheating mechanism predictive and phenomenologically interesting.

\acknowledgments
The work of P.S. and B.S. is supported in part by NSF grant ${\rm PHY}$-${\rm 2014075}$. The work of K.S. is supported in part by DOE Grant desc0009956.


\appendix
\section{Populating a Dark Sector and Its Subsequent Equilibration: General Framework}
\label{sec:DSgeneral}
In this Appendix, we consider the thermal history of a self-interacting dark sector being populated through an unspecified mechanism by relativistic and far from equilibrium DM particles.  We examine the conditions that need to be met for this sector to reach chemical equilibrium. \textit{This Appendix is independent of the initial reheating mechanism, in particular, the physics of PBHs}.

\subsection{Kinetic and Chemical Equilibrium}
We assume that the dark sector content is a real scalar field with the Lagrangian given by Eq.~(\ref{eq:lagrangian}).
We also assume that the dark sector is populated at time $t=\tau$ by relativistic and far from equilibrium DM particles via an unspecified mechanism, and the initial energy density of the dark sector is negligible compared to the energy density of the visible sector.

Kinetic equilibrium requires that the rate of elastic scattering processes becomes comparable to the expansion rate of the Universe. If the initial number density of DM particles is not large enough, thermalization may not happen quickly after production, but it can happen afterwards, as long as DM particles are relativistic. 
Due to the expansion of the Universe, the energy of DM particles decreases and scales inversely with the scale factor until it drops to the DM mass scale which happens at time $t=t_m$, where $t_m>\tau$. After $t_m$, the energy can be replaced by the mass of the particles, $m_\chi$. Therefore, if thermalization does not happen quickly after production, the ratio of rate of elastic scattering processes, $\Gamma_{\chi, 2\rightarrow2}(t)$, to the Hubble expansion rate, $H(t)$, at time $t$ after production and before particles become non-relativistic, is given by:
\begin{equation}
\frac{\Gamma_{\chi, 2\rightarrow2}(t)}{H(t)}\sim\frac{n_\chi(t)/\bar{E}_\chi^2(t)}{H(t)}=\frac{\Gamma_{\chi,2\rightarrow2}(\tau)}{H(\tau)}
\left[\frac{g_{*,S}(T_V(t))}{g_{*,S}(T_V(\tau))}\right]^{1/3}\sqrt{\frac{g_{*,V}(T_V(\tau))}{g_{*,V}(T_V(t))}}\frac{T_V(\tau)}{T_V(t)},~~~~~\tau\lesssim t\lesssim t_m,
\label{eq:rateelastic}
\end{equation}
where $n_\chi(t)$ is the number density of DM particles at time $t$, $\bar{E}_\chi(t)$ is the average energy of particles at time $t$, $T_V(t)$ is the temperature of the Universe at time $t$, and  $H(t)=\sqrt{4\pi^3g_{*,V}(T_V(t))/45}\,T_V^2(t)/M_\text{Pl}$ is the Hubble expansion rate at time $t$. The quantities $g_{*,V}(T_V(t))$ and $g_{*,S}(T_V(t))$ respectively count the total number of relativistic degrees of freedom of the visible sector, and 
the number of relativistic degrees of freedom contributing to the entropy of the visible sector, at time $t$, and are given by
\begin{equation}
 g_{*,V}(T_V)=\sum_B g_B\left(\frac{T_B}{T_V}\right)^4+\frac{7}{8}\sum_F g_F\left(\frac{T_F}{T_V}\right)^4,~~~~~g_{*,S}(T_V)=\sum_B g_B\left(\frac{T_B}{T_V}\right)^3+\frac{7}{8}\sum_F g_F\left(\frac{T_F}{T_V}\right)^3,
\label{eq:gstar}
\end{equation}
where the sum includes all the bosonic ($B$) and fermionic ($F$) degrees of freedom with temperatures equal to $T_B$
and $T_F$ respectively. 

It is clear that if the rate of interactions does not become comparable with the Hubble expansion rate while the particles are relativistic (until $t_m$), it cannot happen after that. The time $t_m$ is easily obtained from
\begin{equation}
T_V(t_m)\simeq\left[\frac{g_{*,S}(T_V(\tau))}{g_{*,S}(T_V(t_m))}\right]^{1/3}\frac{m_\chi}{\bar{E}_\chi(\tau)}T_V(\tau).
\label{eq:tvattm}
\end{equation}

The time at which the dark sector may reach kinetic equilibrium, $t=t_\text{kin-eq}$, where $\tau\lesssim t_\text{kin-eq}\lesssim t_m$, is given by
\begin{equation}
T_V(t_\text{kin-eq})\simeq
\frac{\Gamma_{\chi,2\rightarrow2}(\tau)}{H(\tau)}
\left[\frac{g_{*,S}(T_V(t_\text{kin-eq}))}{g_{*,S}(T_V(\tau))}\right]^{1/3}\sqrt{\frac{g_{*,V}(T_V(\tau))}{g_{*,V}(T_V(t_\text{kin-eq}))}}T_V(\tau).
\label{eq:kineticeqcondition}
\end{equation}

Demanding that kinetic equilibrium is established before DM particles become non-relativistic, i.e.~ $t_\text{kin-eq}\lesssim t_m$, or equivalently, $T_V(t_\text{kin-eq})\gtrsim T_V(t_m)$, leads to a lower bound on the initial abundance of DM particles:
\begin{equation}
    n_\chi(\tau)\gtrsim\left[\frac{g_{*,S}^2(T_V(\tau))}{g_{*,S}(T_V(t_\text{kin-eq}))g_{*,S}(T_V(t_m))}\right]^{1/3} \sqrt{\frac{g_{*,V}(T_V(t_\text{kin-eq}))}{g_{*,V}(T_V(\tau))}}
    m_\chi \bar{E}_\chi(\tau)H(\tau).
    \label{eq:reachkineticeqbeforetm}
\end{equation}

After attaining kinetic equilibrium, the dark sector can be described by a temperature, $T_{\chi}(t_\text{kin-eq})$, and a non-zero chemical potential, $\mu_\chi(t_\text{kin-eq})$. Since kinetic equilibrium happens when particles are still relativistic, dark sector number density and energy density can be expressed as:
\begin{equation}
n_{\chi}(t_\text{kin-eq})=\frac{g_{*,\chi}}{\pi^2}\text{exp}\left[\frac{\mu_\chi(t_\text{kin-eq})}{T_\chi(t_\text{kin-eq})}\right]T_\chi^3(t_\text{kin-eq}), ~~~~~\rho_{\chi}(t_\text{kin-eq})=\frac{3g_{*,\chi}}{\pi^2}\text{exp}\left[\frac{\mu_\chi(t_\text{kin-eq})}{T_\chi(t_\text{kin-eq})}\right]T_{\chi}^4(t_\text{kin-eq}),
\label{eq:densitykineq}
\end{equation}
where $g_{*,\chi}=1$.

The temperature of the dark sector after establishing kinetic equilibrium, $T_\chi(t_\text{kin-eq})$,  can be obtained as 
\begin{equation}
   T_\chi(t_\text{kin-eq})=\frac{1}{3}\frac{\rho_\chi(t_\text{kin-eq})}{n_\chi(t_\text{kin-eq})}=\frac{1}{3}\left[\frac{g_{*,S}(T_V(t_\text{kin-eq}))}{g_{*,S}(T_V(\tau))}\right]^{1/3}\frac{T_V(t_\text{kin-eq})}{T_V(\tau)}\frac{\rho_\chi(\tau)}{n_\chi(\tau)}.
   \label{eq:Tkinteq}
\end{equation}

 After reaching kinetic equilibrium, number-changing processes, $2\rightarrow 3$, can drive the dark sector toward chemical equilibrium, in which case the dark sector can be described only by a temperature. Chemical equilibrium will be established if and only if the rate of the number-changing processes becomes comparable with the Hubble expansion rate. The ratio of these rates can be recast as
 \begin{eqnarray}
   \nonumber\frac{\Gamma_{\chi,2\rightarrow 3}(t)}{H(t)}&\sim&\frac{n_\chi(t)m_\chi^2/T_\chi^4(t)}{H(t)}=\frac{\Gamma_{\chi,2\rightarrow 3}(t_\text{kin-eq})}{H(t_\text{kin-eq})}\left[\frac{g_{*,S}(T_V(t_\text{kin-eq}))}{g_{*,S}(T_V(t))}\right]^{1/3}\sqrt{\frac{g_{*,V}(T_V(t_\text{kin-eq}))}{g_{*,V}(T_V(t))}}
   \left(\frac{T_V(t_\text{kin-eq})}{T_V(t)}\right)^3\\
  \nonumber &=&\frac{\Gamma_{\chi,2\rightarrow 2}(t_\text{kin-eq})}{H(t_\text{kin-eq})}\frac{\Gamma_{\chi,2\rightarrow 3}(t_\text{kin-eq})}{\Gamma_{\chi,2\rightarrow 2}(t_\text{kin-eq})}\left[\frac{g_{*,S}(T_V(t_\text{kin-eq}))}{g_{*,S}(T_V(t))}\right]^{1/3}\sqrt{\frac{g_{*,V}(T_V(t_\text{kin-eq}))}{g_{*,V}(T_V(t))}}
   \left(\frac{T_V(t_\text{kin-eq})}{T_V(t)}\right)^3\\
 &\gtrsim& \left(\frac{m_\chi}{T_\chi(t_\text{kin-eq})}\right)^2\left[\frac{g_{*,S}(T_V(t_\text{kin-eq}))}{g_{*,S}(T_V(t))}\right]^{1/3}\sqrt{\frac{g_{*,V}(T_V(t_\text{kin-eq}))}{g_{*,V}(T_V(t))}}
   \left(\frac{T_V(t_\text{kin-eq})}{T_V(t)}\right)^3,~~~~ t_\text{kin-eq}\lesssim t \lesssim t_m,
 \label{eq:2to3rate}
 \end{eqnarray}
 
and since during the time interval, $t_\text{kin-eq}\lesssim t \lesssim t_m$, we have
\begin{equation}
\frac{T_V(t_\text{kin-eq})}{T_V(t)}\lesssim \frac{T_V(t_\text{kin-eq})}{T_V(t_m)}=\left[\frac{g_{*,S}(T_V(t_m))}{g_{*,S}(T_V(t_\text{kin-eq}))}\right]^{1/3}\frac{T_\chi(t_\text{kin-eq})}{T_\chi(t_m)}\simeq\left[\frac{g_{*,S}(T_V(t_m))}{g_{*,S}(T_V(t_\text{kin-eq}))}\right]^{1/3}\frac{3T_\chi(t_\text{kin-eq})}{m_\chi},
\end{equation}

 the right-hand side of Eq.~(\ref{eq:2to3rate}) definitely exceeds one at some time after establishment of kinetic equilibrium and before particles become non-relativistic. This shows that if particles are abundant enough to reach kinetic equilibrium, they will also reach chemical equilibrium afterwards. 
 
 Therefore, the time at which the dark sector reaches chemical equilibrium, $t=t_\text{chem-eq}$, where $t_\text{kin-eq}\lesssim t_\text{chem-eq}\lesssim t_m$, is estimated as
 \begin{equation}
      T_V(t_\text{chem-eq})=\left[\frac{g_{*,S}(T_V(t_\text{kin-eq}))}{g_{*,S}(T_V(t_\text{chem-eq}))}\right]^{1/9}\left[\frac{g_{*,V}(T_V(t_\text{kin-eq}))}{g_{*,V}(T_V(t_\text{chem-eq}))}\right]^{1/6}\left(\frac{m_\chi}{T_\chi(t_\text{kin-eq})}\right)^{2/3}T_V(t_\text{kin-eq}),
     \label{eq:chemeqcondition}
 \end{equation}
 where $T_V(t_\text{kin-eq})$ and $T_\chi(t_\text{kin-eq})$ are given by Eqs.~(\ref{eq:kineticeqcondition}) and~(\ref{eq:Tkinteq}) respectively.

\subsection{Chemical Equilibrium, Relativistic Gas}
If dark sector particles reach chemical equilibrium while they are still relativistic, then dark sector number density and energy density can be expressed as:
\begin{equation}
   n_{\chi, \text{rel}}(t_\text{chem-eq})=\frac{\zeta(3)}{\pi^2}g_{*,\chi}T_{\chi,\text{rel}}^3(t_\text{chem-eq}), ~~~~~\rho_{\chi,\text{rel}}(t_\text{chem-eq})=\frac{\pi^2}{30}g_{*,\chi}T_{\chi,\text{rel}}^4(t_\text{chem-eq}).
   \label{eq:densitychemrel}
\end{equation}
The energy density of the dark sector at equilibrium time can be obtained by redshifting the energy density at production time to the equilibrium time as
\begin{equation}
    \rho_{\chi}(t_\text{chem-eq})=\rho_{\chi}(\tau)\left[\frac{g_{*,S}(T_V(t_\text{chem-eq}))}{g_{*,S}(T_V(\tau))}\right]^{4/3}\left(\frac{T_V(t_\text{chem-eq})}{T_V(\tau)}\right)^4.
    \label{eq:gettempfromenergyrel}
\end{equation}

Therefore the temperature of the dark sector after chemical equilibrium is given by
\begin{equation}
    T_{\chi,\text{rel}}(t_\text{chem-eq})=\left(\frac{30}{\pi^2}\frac{1}{g_{*,\chi}}\right)^{1/4}\left[\frac{g_{*,S}(T_V(t_\text{chem-eq}))}{g_{*,S}(T_V(\tau))}\right]^{1/3}\frac{T_V(t_\text{chem-eq})}{T_V(\tau)}\rho_\chi^{1/4}(\tau).
    \label{eq:TXchemrel}
\end{equation}

This estimate of temperature is consistent as long as $T_{\chi,\text{rel}}(t_\text{eq})\gtrsim m_\chi/3$, which leads to a lower bound on initial abundance of DM.

The ratio of the dark sector temperature to the temperature of the visible sector at equilibrium time, $\xi$, is obtained as
\begin{equation}
\xi=\frac{T_{\chi,\text{rel}}(t_\text{chem-eq})}{T_V(t_\text{chem-eq})}=\left(\frac{30}{\pi^2}\frac{1}{g_{*,\chi}}\right)^{1/4}\left[\frac{g_{*,S}(T_V(t_\text{chem-eq}))}{g_{*,S}(T_V(\tau))}\right]^{1/3}\frac{\rho_\chi^{1/4}(\tau)}{T_V(\tau)}.
\label{eq:ratioTXchemreltoTV}
\end{equation}
Since we assume that the visible sector always dominates the energy density of the Universe, therefore Eq.~(\ref{eq:ratioTXchemreltoTV}) shows that if the dark sector reaches chemical equilibrium while DM particles are still relativistic, its equilibrium temperature is always colder than the temperature of the visible sector at that time.

To check the possibility of a cannibal phase at $t=t_m$, when DM particles become non-relativistic, i.e.~ $T_\chi(t_m)\sim m_\chi/3$, we need to evaluate the Hubble expansion rate at that time, which depends on the temperature of the visible sector at $t=t_m$.
Since from $t_\text{eq}$ to $t_m$, dark sector is in chemical equilibrium, its total entropy is conserved. Conservation of total entropy in each sector separately determines the evolution of the temperature of each sector and therefore temperature of the visible sector at $t_m$ can be evaluated from
\begin{equation}
\frac{T_{\chi,\text{rel}}(t_\text{chem-eq})}{T_\chi(t_m)}=\left[\frac{g_{*,S}(T_V(t_\text{chem-eq}))}{g_{*,S}(T_V(t_{m}))}\right]^{1/3}\frac{T_V(t_\text{chem-eq})}{T_V(t_m)}.
\label{eq:Tvattmaftereq}
\end{equation}
At $t=t_m$, we have
\begin{equation}
\frac{\Gamma_{\chi,3\rightarrow 2}(t_m)}{H(t_m)}\sim\frac{n_\chi^2(t_m)/m_\chi^5}{H(t_m)},
\label{eq:3to2ratetoHRel}
\end{equation}
where $n_\chi(t_m)\sim\frac{g_{*,\chi}}{e^{3}(6\pi)^{3/2}} m_\chi^3$ and $H(t_m)=H(T_V(t_m))$ with $T_V(t_m)$ obtained from Eq.~(\ref{eq:Tvattmaftereq}).

If $\Gamma_{\chi,3\rightarrow 2}(t_m)/H(t_m)<1$, DM particles decouple after becoming non-relativistic ($t_\text{dec}=t_m$) and temperature of the dark sector at $t_m$ sets the relic abundance of the dark sector. This condition gives rise to an upper limit on the initial abundance of the DM.
In this case, after $t_m$, temperature of the dark sector scales as the temperature of a decoupled non-relativistic particle and dark sector also develops a chemical potential~\cite{Kolb:1990vq}:
\begin{equation}
T_\chi(t)= T_\chi(t_\text{dec})\left(\frac{a(t_\text{dec})}{a(t)}\right)^2,~~~~~\mu_\chi(t)= m_\chi\left(1-\frac{T_\chi(t)}{T_\chi(t_\text{dec})}\right).
\label{eq:TXafterdec}
\end{equation}

The amount of DM today is given by
\begin{equation}
 Y_\chi =\frac{n_\chi(t_0)}{s(t_0)}=\frac{n_\chi(t_\text{dec})}{s(t_\text{dec})}=\frac{g_{*,\chi}\left(\frac{m_\chi T_\chi(t_\text{dec})}{2\pi}\right)^{3/2}e^{-m_\chi/T_\chi(t_\text{dec})}}{s(T_V(t_\text{dec}))},
 \label{eq:YXaftertdecRDcannLrel}
\end{equation}

where a subscript $0$ means the quantity is evaluated today, $s(t)$ is the entropy density given by
\begin{equation}
s(t)=s(T_V(t))=\frac{2\pi^2}{45}g_{*,S}(T_V(t))T_V^3(t).
\label{eq:entropy}
\end{equation}

Given the amount of DM today, the relic abundance of DM today can be obtained from 
\begin{equation}
\Omega_\chi=\frac{\rho_{\chi,0}}{\rho_c}=\frac{m_\chi Y_\chi }{\rho_c}s(t_0),
\label{eq:relicdef}
\end{equation}

with the critical energy density of the Universe, $\rho_c$, and entropy density of the Universe today, $s(t_0)$, equal to~\cite{Aghanim:2018eyx}:
\begin{equation}
   \rho_c=1.0537\times 10^{-5}\, h^2\,\,\text{GeV}\text{ cm}^{-3},~~~~~s(t_0)=2891.2\left(\frac{T_0}{2.7255}\right)^3 \text{cm}^{-3}.
   \label{eq:constants}
\end{equation}
Although precise evaluation of DM relic abundance demands solving Boltzmann equation, for simplicity, in this work we estimate it by assuming instantaneous freeze-out of DM.

On the other hand, when $\Gamma_{\chi,3\rightarrow 2}(t_m)/H(t_m)\gtrsim1$,
dark sector enters a cannibal phase.
During the cannibal phase, since the non-relativistic dark sector is still in chemical equilibrium, the entropy of the dark sector is conserved and therefore we have
\begin{equation}
 a^3 s_\chi=a^3\frac{\rho_\chi+p_\chi}{T_\chi}\simeq a^3\frac{\rho_\chi}{T_\chi}\simeq a^3\frac{m_\chi n_\chi}{T_\chi}=a^3 g_{*,\chi}\frac{m_\chi}{T_\chi}\left(\frac{m_\chi T_\chi}{2\pi}\right)^{3/2} e^{-m_\chi/T_\chi}=\text{const.},
 \label{eq:cannibalentropyrel}
\end{equation}
where we used the fact that $p_\chi\simeq n_\chi T_\chi \ll \rho_\chi$.
The entropy of the dark sector at $t_m$ when $T_\chi\sim m_\chi/3$, determines the const. on the right hand side of Eq.~(\ref{eq:cannibalentropyrel}) which makes it possible to obtain $T_\chi(t)$ during the cannibal phase as 
\begin{equation}
T_\chi (t)=\frac{2m_\chi}{W\left[6\,e^6\left(\frac{a(t)}{a(t_m)}\right)^6\right]}=\frac{2m_\chi}{W\left[6\,e^6\left(\frac{g_{*,S}(T_V(t_m))}{g_{*,S}(T_V(t))}\right)^2\left(\frac{T_V(t_m)}{T_V(t)}\right)^6\right]},
 \label{eq:TXduringcannibalrel}
\end{equation}

where Lambert $W$- function is defined to be the function satisfying
$W(x) e^{W(x)}=x$ and can be approximated as $W(x)\sim x$ when $x\lesssim 1$ and $W(x)\sim \text{ln}\,x-\text{ln}\,\text{ln}\,x$ when $x\gtrsim 1$, and $T_V(t_m)$ is given by Eq.~~(\ref{eq:Tvattmaftereq}).

The logarithmic decrease of dark sector temperature with the scale factor~\cite{Carlson:1992fn}, is the key characteristic of the cannibal phase which is milder than the power-law decrease of the temperature of a decoupled non-relativistic sector. 

Cannibal phase continues until $t=t_\text{dec}$, when the rate of number-changing process becomes smaller than the Hubble expansion rate. Therefore the time of decoupling, $t_\text{dec}$, can be estimated from
\begin{equation}
\frac{\Gamma_{\chi,3\rightarrow 2}(t_\text{dec})}{H(t_\text{dec})}\sim\frac{n_\chi^2(t_\text{dec})/m_\chi^5}{H(t_\text{dec})}\simeq 1,
 \label{eq:decouplingcondrelcann}
\end{equation}
where $n_\chi(t_\text{dec})=n_\chi(T_\chi(t_\text{dec}))$ and $H(t_\text{dec})=H(T_V(t_\text{dec}))$. By using Eqs.~(\ref{eq:TXduringcannibalrel}) and~(\ref{eq:decouplingcondrelcann}), we obtain the time of decoupling, $t_\text{dec}$, as

\begin{eqnarray}
    \nonumber T_V(t_\text{dec})&\simeq&\frac{2\sqrt{2}e^{3/2}\pi^{9/8}}{5^{1/8}}\frac{g_{*,V}^{1/8}(T_V(t_\text{dec}))}{\sqrt{g_{*,\chi}}}\left(\frac{g_{*,S}(T_V(t_m))}{g_{*,S}(T_V(t_\text{dec}))}\right)^{1/2}\left(\frac{T_V^6(t_m)}{M_\text{Pl} m_\chi}\right)^{1/4}\\
   &\times&W\left[\frac{3^{1/4}5^{3/16}}{4\sqrt{2}e^{3/4}\pi^{27/16}}\left(\frac{g_{*,\chi}^4}{g_{*,V}(T_V(t_\text{dec}))}\right)^{3/16}\left(\frac{g_{*,S}(T_V(t_\text{dec}))}{g_{*,S}(T_V(t_m))}\right)^{1/4}\left(\frac{M_\text{Pl} m_\chi}{T_V(t_m)^2}\right)^{3/8}\right]^{1/2}.
    \label{eq:TVtdecoupling}
\end{eqnarray}

After $t_\text{dec}$, DM particles decouple and dark sector is characterized by a temperature and a chemical potential given by Eq.~(\ref{eq:TXafterdec}) where 
\begin{equation}
   T_\chi(t_\text{dec})= \frac{2m_\chi}{W\left[6\,e^6\left(\frac{g_{*,S}(T_V(t_m))}{g_{*,S}(T_V(t_\text{dec}))}\right)^2\left(\frac{T_V(t_m)}{T_V(t_\text{dec})}\right)^6\right]}.
 \label{eq:TXdecrel}
\end{equation}
The amount of DM today is given by Eq.~(\ref{eq:YXaftertdecRDcannLrel}).

\subsection{Chemical Equilibrium, Non-relativistic Gas}
If self-interaction leads the dark sector to reach chemical equilibrium when DM particles are non-relativistic, then number density and energy density of DM particles can be expressed as:
\begin{eqnarray}
\nonumber &n_{\chi, \text{non-rel}}(t_\text{chem-eq})&=g_{*,\chi}\left(\frac{m_\chi T_{\chi,\text{non-rel}}(t_\text{chem-eq})}{2\pi}\right)^{3/2}\text{exp}\left[\frac{-m_\chi}{T_{\chi,\text{non-rel}}(t_\text{chem-eq})}\right],\\ 
&\rho_{\chi, \text{non-rel}}(t_\text{chem-eq})&=m_\chi n_{\chi, \text{non-rel}}(t_\text{chem-eq}).
\label{eq:densitychemnonrel}
\end{eqnarray}

Temperature of the dark sector is obtained from Eq.~(\ref{eq:densitychemnonrel}) as 
\begin{eqnarray}
T_{\chi,\text{non-rel}}(t_\text{chem-eq})=\frac{2m_\chi}{3W\left[\frac{g_{*,\chi}^{2/3}}{3\pi}\left(\frac{m_\chi^4}{\rho_\chi(t_\text{chem-eq})}\right)^{2/3}\right]}.
\label{eq:TXchemnonrel}
\end{eqnarray}
where $\rho_\chi(t_\text{chem-eq})$ is given by Eq.~(\ref{eq:gettempfromenergyrel}).
This calculation of temperature is consistent as long as $T_{\chi,\text{non-rel}}(t_\text{chem-eq})\lesssim m_\chi/3$.

The possibility of a cannibal phase can be explored by estimating the ratio of the rate of number-changing process to the Hubble expansion rate at $t_\text{chem-eq}$, given by 
\begin{equation}
\frac{\Gamma_{\chi,3\rightarrow 2}(t_\text{chem-eq})}{H(t_\text{chem-eq})}\sim\frac{n_\chi^2(t_\text{chem-eq})/m_\chi^5}{H(t_\text{chem-eq})}=\frac{\rho_\chi^2(t_\text{chem-eq})/m_\chi^7}{H(t_\text{chem-eq})}.
\label{eq:3to2ratetoHnonRel}
\end{equation}

If $\Gamma_{\chi,3\rightarrow 2}(t_\text{chem-eq})/H(t_\text{chem-eq})<1$, DM particles decouple quickly after reaching to chemical equilibrium ($t_\text{dec}\simeq t_\text{chem-eq}$) and temperature of the dark sector at $t_\text{chem-eq}$ sets the relic abundance of the dark sector. This condition is equivalent to an upper limit on the initial abundance of the DM. After $t_\text{chem-eq}$, dark sector is described by a temperature and a chemical potential given by Eq.~(\ref{eq:TXafterdec}), and the amount of DM today is also given by Eq.~(\ref{eq:YXaftertdecRDcannLrel}) where $t_\text{dec}= t_\text{chem-eq}$.

On the other hand, when $\Gamma_{\chi,3\rightarrow 2}(t_\text{chem-eq})/H(t_\text{chem-eq})\gtrsim1$,
dark sector enters a cannibal phase. 
During the cannibal phase the temperature of the dark sector can be obtained by using conservation of entropy in the dark sector, i.e.~ Eq.~(\ref{eq:cannibalentropyrel}). The entropy of the dark sector at $t_\text{chem-eq}$ when $T_\chi=T_\chi(t_\text{chem-eq})$, determines the const. on the right-hand side of Eq.~(\ref{eq:cannibalentropyrel}) which gives the temperature of the dark sector during the cannibal phase as  
\begin{equation}
    T_\chi(t)=\frac{2m_\chi}{W\left[\frac{1}{4\pi^3}g_{*,\chi}^2
    \left(\frac{g_{*,S}(T_V(t_\text{chem-eq}))}{g_{*,S}(T_V(t))}\right)^2\left(\frac{T_V(t_\text{chem-eq})}{T_V(t)}\right)^6\frac{m_\chi^6}{s_\chi^2(t_\text{chem-eq})}\right]},
    \label{eq:TXduringcannibalnonrel}
\end{equation}

where $s_\chi(t_\text{chem-eq})=\rho_\chi(t_\text{chem-eq})/T_{\chi,\text{non-rel}}(t_\text{chem-eq})$.

Cannibal phase continues until $t_\text{dec}$, when the rate of number-changing process becomes smaller than the Hubble rate. Therefore the time of decoupling can be obtained from Eq.~(\ref{eq:decouplingcondrelcann}) as
\begin{eqnarray}
\nonumber T_V(t_\text{dec})&=&\frac{2^{3/4}\pi^{3/8}}{3^{1/4}5^{1/8}}
g_{*,V}^{1/8}(T_V(t_\text{dec}))\left(\frac{g_{*,S}(T_V(t_\text{chem-eq}))}{g_{*,S}(T_V(t_\text{dec}))}\right)^{1/2}\left(\frac{m_\chi^5 T_V(t_\text{chem-eq})^6}{M_\text{Pl} s_\chi^2(t_\text{chem-eq})}\right)^{1/4}\\
&\times& W\left[\frac{3^{3/8}5^{3/16}}{4\times2^{1/8}\pi^{21/16}}\frac{\sqrt{g_{*,\chi}}}{g_{*,V}^{3/16}(T_V(t_\text{dec}))}\left(\frac{g_{*,S}(T_V(t_\text{dec}))}{g_{*,S}(T_V(t_\text{chem-eq}))}\right)^{1/4}
\left(\frac{M_\text{Pl}^3 s_\chi^2(t_\text{chem-eq})}{m_\chi^3 T_V(t_\text{chem-eq})^6}\right)^{1/8}\right]^{1/2}.
\label{eq:decouplingcondnonrelcann}
\end{eqnarray}
After $t_\text{dec}$, dark sector is represented by a temperature and a chemical potential given by Eq.~(\ref{eq:TXafterdec}) where
\begin{equation}
    T_\chi(t_\text{dec})=\frac{2m_\chi}{W\left[\frac{1}{4\pi^3}g_{*,\chi}^2
    \left(\frac{g_{*,S}(T_V(t_\text{chem-eq}))}{g_{*,S}(T_V(t_\text{dec}))}\right)^2\left(\frac{T_V(t_\text{chem-eq})}{T_V(t_\text{dec})}\right)^6\frac{m_\chi^6}{s_\chi^2(t_\text{chem-eq})}\right]},
    \label{eq:TXdeccannibalnonrel}
\end{equation}
and the amount of DM today is obtained from Eq.~(\ref{eq:YXaftertdecRDcannLrel}).

\section{Primordial Black Holes, Formation and Evaporation}
\label{sec:PBH}
In this Appendix, we review the formation of PBHs and their subsequent evaporation by following Ref.~\cite{Gondolo:2020uqv}.

In the early Universe, density fluctuations, $\delta\rho/\rho$, which grow after entering the
horizon, can collapse into a PBH if they are  greater than the equation of state parameter, $w\equiv p/\rho$.  
The overdense region can overcome the pressure of the radiation, if its size is larger than the Jeans length, which is $\sqrt{w}$ times the horizon size~\cite{Carr:1975qj, Carr:1974nx}. The mass of the PBH, which is bounded by the total mass within the horizon~\cite{Carr:1975qj}, for formation during radiation-dominated epoch can be expressed as
\begin{equation}
M_i=\frac{4\pi}{3}\gamma \rho_V(T_V(t_i))  H^{-3}(T_V(t_i)),
\label{eq:initialmass}
\end{equation}
where $\gamma\sim w^{3/2}\approx0.2$ in a radiation-dominated Universe, $T_V(t_i)$ is the temperature of the Universe at  PBH formation time, $t_i$, and $\rho_V(T)$ is the energy density of the Universe, given by
\begin{equation}
\rho_V(T)=\frac{\pi^2}{30}g_{*,V}(T)T^4.
\label{eq:energydenUniverse}
\end{equation} 
The initial mass of PBHs is related to their time of formation by 
Eq.~(\ref{eq:initialmass}).

By emitting all the particles in the spectrum which are lighter than its temperature, a black hole loses its mass through Hawking evaporation~\cite{Hawking:1974sw}. Since after ignoring greybody factors, the Hawking radiation can be described as a black body radiation, the energy spectrum of the $i$th emitted species by a non-rotating black hole with zero charge is given by
\begin{equation}
\frac{d^2u_i(E,t)}{dtdE}=\frac{g_i}{8\pi^2}\frac{E^3}{e^{E/T_\text{BH}}\pm1} ,
\label{eq:HWrate}
\end{equation}
($+$ for fermion emission and $-$ for boson emission) where $u_i(E,t)$ is the total radiated energy per unit area, $g_i$ counts the number of degrees of freedom of the $i$th species, $E$ is the energy of the emitted particle, and $T_\text{BH}$ is the horizon temperature of the black hole given by
\begin{equation}
T_\text{BH}=\frac{M_\text{Pl}^2}{8\pi M_\text{BH}}.
\label{eq:BHtemp}
\end{equation}
The mass loss rate of a black hole due to Hawking evaporation is obtained from Eqs.~(\ref{eq:HWrate}) and (\ref{eq:BHtemp}) as
\begin{equation}
\frac{dM_\text{BH}}{dt}=-4\pi r_\text{S}^2\sum_i\int_0^\infty  \frac{d^2u_i(E,t)}{dtdE} dE=-\frac{g_*(T_\text{BH})}{30720 \pi}\frac{M_\text{Pl}^4}{M_\text{BH}^2},
\label{eq:massloss}
\end{equation}
where $g_*(T_\text{BH})$ counts the total number of relativistic degrees of freedom emitted by the black hole, and $r_\text{S}=2 M_\text{BH}/M_\text{Pl}^2$ is the Schwarzschild radius 
of the black hole.
The time evolution of the mass of a black hole with initial mass $M_i$ formed at $t_i$ is evaluated by integrating Eq.~(\ref{eq:massloss}),
\begin{equation}
M(t)=M_i\left(1-\frac{t-t_i}{\tau}\right)^{1/3},
\label{eq:massBH}
\end{equation}
where
\begin{equation}
\tau=\frac{10240\pi}{g_*(T_\text{BH})}\frac{M_i^3}{M_\text{Pl}^4},
\label{eq:tauBH}
\end{equation}
is the lifetime of black hole.

Eq.~(\ref{eq:initialmass}) can be recast to give the temperature of the Universe at the time of formation of PBHs, $T_V(t_i)$, as
\begin{equation}
T_V(t_i)=\frac{\sqrt{3}\,5^{1/4}\gamma^{1/2}}{2\pi^{3/4}g_{*,V}^{1/4}(t_i)}\left(\frac{M_\text{Pl}^{3}}{M_\text{BH}}\right)^{1/2}.
\label{eq:TVi}
\end{equation}

The temperature of the Universe at the time of evaporation of PBHs, which defined as $T_V (t_i+\tau)\simeq T_V(\tau)$, in a radiation-dominated epoch where $H(t)=1/(2t)$, is calculated as
\begin{equation}
T_V(\tau)=\frac{\sqrt{3}\,g_{*,V}^{1/4}(\tau)}{64\sqrt{2}\,5^{1/4}\pi^{5/4}}\left(\frac{M_\text{Pl}^{5}}{M_\text{BH}^{3}}\right)^{1/2},
\label{eq:TVtauRD}
\end{equation}
where the Friedmann equation, $H^2=8\pi\rho/3M_\text{Pl}^2$, is used.

The rate of emission of the $i$th emitted species per energy interval can be expressed in terms of its energy spectrum as
\begin{equation}
\frac{d^2N_i}{dtdE}=\frac{4\pi r_\text{S}^2}{E}\frac{d^2u_i}{dtdE}=\frac{g_i}{2\pi}\frac{r_\text{S}^2E^2}{e^{E/T_\text{BH}}\pm1} .
\label{eq:numrate}
\end{equation}
The total number of particles of the $i$th species emitted over the lifetime of the black hole is obtained by integrating Eq.~(\ref{eq:numrate}) over energy and time. For bosons,
\begin{eqnarray}
\nonumber N_i&=&\frac{120\,\zeta(3)}{\pi^3}\frac{g_i}{g_*(T_\text{BH})}\frac{M_\text{BH}^2}{M_\text{Pl}^2},~~~~~~T_\text{BH}>m_i , \\
N_i&=&\frac{15\,\zeta(3)}{8\pi^5}\frac{g_i}{g_*(T_\text{BH})}\frac{M_\text{Pl}^2}{m_i^2},~~~~~~~T_\text{BH}<m_i.
\label{eq:numberemitted}
\end{eqnarray}
The total number of fermionic species is $N_F=\frac{3}{4}\frac{g_F}{g_B}N_B$.

The average energy of the produced particles,
\begin{equation}
\bar{E}=\frac{1}{N_i}\iint\,dEdt\, E\frac{d^2N_i}{dtdE},
\end{equation}
is given by:
\begin{eqnarray}
\nonumber \bar{E}_i&=&\frac{\pi^4}{15\zeta(3)}T_\text{BH},~~~~~~T_\text{BH}>m_i , \\
\bar{E}_i&=&\frac{\pi^4}{15\zeta(3)}m_i,~~~~~~~T_\text{BH}<m_i.
\label{eq:avee}
\end{eqnarray}
To represent the initial abundance of PBHs, it is customary to introduce the dimensionless parameter $\beta$, defined as the initial energy density of PBHs normalized to the radiation energy density:
\begin{equation}
\beta=M_\text{BH}\frac{n_\text{BH}(t_i)}{\rho_\text{V}(t_i)}.
\label{eq:beta}
\end{equation}
Since the energy density of PBHs redshifts as matter, an initially radiation-dominated Universe can eventually become matter-dominated prior to evaporation of PBHs. 
The critical initial abundance of PBHs, $\beta_\text{crit}$, that gives rise to an early matter-dominated era can be obtained by requiring that PBH evaporation happens after an early equality time $t_\text{early-eq}$, at which $\rho_\text{PBH}(t_\text{early-eq})/\rho_\text{rad}(t_\text{early-eq})\sim 1$.
This early equality time is expressed in terms of $T_V(t_i)$ and $\beta_\text{crit}$ as
\begin{equation}
\frac{\rho_\text{PBH}(t_\text{early-eq})}{\rho_\text{rad}(t_\text{early-eq})}=\frac{\rho_\text{PBH}(T_V(t_i))}{\rho_\text{rad}(T_V(t_i))}\frac{T_V(t_i)}{T_V(t_\text{early-eq})}=\beta_\text{crit}\frac{T_V(t_i)}{T_V(t_\text{early-eq})}\sim 1.
\label{eq:conditionmatterdomination}
\end{equation}
An early matter-dominated era is assured provided that $t_\text{early-eq}\lesssim t_\text{eva}$, or equivalently when
\begin{equation}
\beta\geq\beta_\text{crit}=\frac{T_V(\tau)}{T_V(t_i)}=\frac{1}{32\sqrt{10\pi}}\sqrt{g_{*,V}}\frac{1}{\sqrt{\gamma}}\frac{M_\text{Pl}}{M_\text{BH}}.
\label{eq:betac}
\end{equation}
\section{Populating a Dark Sector by PBHs and Its Subsequent Equilibration}
\label{sec:DS}
 In this Appendix, we apply general formalism presented in Appendix~\ref{sec:DSgeneral} to the case where Hawking evaporation of a population of PBHs with a monochromatic mass function in early Universe is responsible for populating a dark sector. 
 
 The initial abundance of PBHs determines if they dominate the energy density of the Universe prior to their evaporation or not;
 $\beta<\beta_\text{crit}\, (\beta>\beta_\text{crit})$ leads to an early radiation (matter)-dominated Universe. In either case, DM mass can be less than or larger than the initial temperature of PBHs which in turn, affects the initial number density and energy density of the emitted particles. The produced DM particles by PBHs may reach chemical equilibrium while they are still relativistic or non-relativistic. They may or may not go through a cannibal phase before decoupling. The possibility of each of these outcomes and their effects on relic abundance of DM today is examined in this Appendix. For simplicity, the time evolution of $g_{*,V}$ and $g_{*,S}$ is not considered here ($g_{*,V}\simeq 106.8$). It does not affect our main results and can be easily recovered by following Appendix~\ref{sec:DSgeneral}.
 
\subsection{Radiation-dominated Universe \boldmath{$(\beta< \beta_\text{crit})$}}
According to general setup discussed in  Appendix~\ref{sec:DSgeneral}, to understand the thermal history of the populated dark sector, we need the number density and energy density of the dark sector at population time. Provided that the evaporation of PBHs happens instantaneously which is a valid assumption to make (see Eq.~(\ref{eq:massBH})), then the fraction of energy density of PBHs which transfers into the dark sector is given by 
\begin{equation}
\rho_{\chi,\text{BH}}(\tau)=n_\text{BH}(\tau)N_\chi\bar{E}_\chi=\frac{\pi^2}{30}g_{*,V}\beta N_\chi\frac{\bar{E}_\chi}{M_\text{BH}}T_V(t_i)T_V^3(\tau),
\label{eq:energyinjectedtoDSRD}
\end{equation}

and the number density of bosonic DM particles after PBHs evaporation can be evaluated as
\begin{equation}
n_\chi(\tau)=N_\chi n_\text{BH}(\tau)=\frac{\pi^2}{30}g_{*,V}\beta N_\chi\frac{T_V(t_i)T_V^3(\tau)}{M_\text{BH}}.
\label{eq:numberdensityinjectedtoDSRD}
\end{equation}

By using Eqs.(\ref{eq:numberemitted}) and (\ref{eq:avee}), the energy density of the dark sector immediately after PBHs evaporation can be recast as
\begin{eqnarray}
\nonumber \rho_{\chi, \text{BH}}(\tau)&=&\frac{\pi^2}{30}g_{*,\chi}\beta T_V(t_i)T_V(\tau)^3 ,~~~~~~T_\text{BH}>m_\chi , \\
\rho_{\chi,\text{BH}}(\tau)&=&\frac{\pi}{240}g_{*,\chi}\beta \frac{M_\text{Pl}^2}{M_\text{BH}m_\chi}T_V(t_i)T_V^3(\tau)=\frac{\pi^2}{30}g_{*,\chi}\beta\frac{T_\text{BH}}{m_\chi} T_V(t_i)T_V^3(\tau),~~~~~~~T_\text{BH}<m_\chi.
\label{eq:energyinjectedtoDSRD2}
\end{eqnarray}

It is worth mentioning that the amount of energy transferred into the dark sector when DM is heavier than the initial temperature of PBHs, is suppressed by the factor $T_\text{BH}/m_\chi$ with respect to the case where DM is lighter than the initial temperature of PBHs.\footnote{We notice that our result is not in agreement with the one reported in Ref.~\cite{Bernal:2020kse} which is wrongly proportional to $m_\chi/T_\text{BH}$.}

\subsubsection{\boldmath{$m_\chi < T_\text{BH}$}}
\textbf{Kinetic and Chemical Equilibrium:} It can easily be shown that immediately after evaporation, the rate of the elastic scattering processes is smaller than the Hubble expansion rate and therefore kinetic equilibration cannot happen instantaneously. By using Eqs.~(\ref{eq:rateelastic}), (\ref{eq:TVtauRD}), (\ref{eq:avee}) and (\ref{eq:numberdensityinjectedtoDSRD}) we have:
\begin{equation}
\frac{\Gamma_{\chi, 2\rightarrow2}(\tau)}{H(\tau)}\sim\frac{2025\zeta^3(3)}{64\pi^{11}}g_{*,\chi}\frac{\beta}{\beta_\text{crit}}<\frac{2025\zeta^3(3)}{64\pi^{11}}g_{*,\chi}\simeq2\times10^{-4}g_{*,\chi},
\label{ratekinRDtauLrel}
\end{equation}
since we assume a radiation-dominated Universe, i.e.~$\beta<\beta_\text{crit}$. Although to evaluate the rate of elastic scattering processes in Eq.~(\ref{ratekinRDtauLrel}), self-coupling is assumed to be $\lambda\sim1$, but even saturating the perturbative unitarity limit cannot change this result. The reason is that the elastic scattering cross-section in Eq.~(\ref{ratekinRDtauLrel}) is estimated as $\sigma_{2\rightarrow2}=1/E^2$, which can be modified by multiplying by $\lambda^4/(32 \pi)$ to include the effect of the self-coupling.
Therefore, a self-coupling as large as $\lambda\sim 4\pi$, cannot make the rate of elastic scattering processes comparable with Hubble expansion rate at evaporation time.

Kinetic equilibrium may happen later at $t=t_\text{kin-eq}$, when the rate of elastic scattering processes becomes comparable to the Hubble expansion rate. $t_\text{kin-eq}$ is obtained from Eq.~(\ref{eq:kineticeqcondition}) as
\begin{equation}
T_V(t_\text{kin-eq})\sim\frac{2025\sqrt{5}\zeta^3(3)}{\sqrt{2}\pi^{21/2}}\frac{g_{*,\chi}}{g_{*,V}^{1/2}}\sqrt{\gamma}\beta\frac{M_\text{BH}}{M_\text{Pl}}T_V(\tau),
\label{eq:ratekineticRDEQLrel}
\end{equation}

and this should happen before $t_m$, i.e.~, $T_V(t_\text{kin-eq})\gtrsim T_V(t_m)$, which by using Eq.~(\ref{eq:tvattm}) leads to a lower bound on the initial abundance of PBHs, $\beta_\text{kin}$,
\begin{equation}
    \beta\gtrsim\beta_\text{kin}\equiv\frac{8\sqrt{2}\pi^{15/2}}{135\sqrt{5}\zeta^2(3)}\frac{\sqrt{g_{*,V}}}{g_{*,\chi}}\frac{1}{\sqrt{\gamma}}\frac{m_\chi}{M_\text{Pl}}.
    \label{eq:betakin}
\end{equation}

By using Eqs.~(\ref{eq:Tkinteq}), (\ref{eq:TVtauRD}), (\ref{eq:avee}) and (\ref{eq:numberdensityinjectedtoDSRD}) we can estimate the temperature of the dark sector after reaching kinetic equilibrium at $t_\text{kin-eq}$ as:
\begin{equation}
T_{\chi}(t_\text{kin-eq})=\frac{45\sqrt{5}\zeta^2(3)}{8\sqrt{2}\pi^{15/2}}\frac{g_{*,\chi}}{\sqrt{g_{*,V}}}\sqrt{\gamma}\beta M_\text{Pl}.
\label{eq:TkintRDLrel}
\end{equation}

From Eq.~(\ref{eq:chemeqcondition}) we can find the time at which the dark sector reaches chemical equilibrium, $t_\text{chem-eq}$ as
\begin{equation}
T_V(t_\text{chem-eq})= 
\frac{135\times3^{1/6}5^{1/4}\zeta^{5/3}(3)}{16\times 2^{2/3}\pi^{27/4}}g_{*,V}^{1/12}g_{*,\chi}^{1/3}\gamma^{1/6}\beta^{1/3}\left(\frac{m_\chi^4 M_\text{Pl}^5}{M_\text{BH}^3}\right)^{1/6}.
\label{eq:TVtchemRD}
\end{equation}

\textbf{Chemical Equilibrium, Relativistic Gas:} If the dark sector attains chemical equilibrium while DM particles are still relativistic, the temperature of the dark sector can be obtained by using Eqs.~(\ref{eq:TXchemrel}), (\ref{eq:TVtauRD}), (\ref{eq:avee}),  (\ref{eq:numberdensityinjectedtoDSRD}), and (\ref{eq:energyinjectedtoDSRD2}) as:
\begin{equation}
T_{\chi,\text{rel}}(t_\text{chem-eq})=\frac{135\times3^{1/6}5^{3/8}\zeta^{5/3}(3)}{8\times 2^{7/24}\pi^{53/8}}\frac{g_{*,\chi}^{1/3}}{g_{*,V}^{1/24}}\gamma^{7/24}\beta^{7/12}\left(\frac{m_\chi^8M_\text{Pl}^7}{M_\text{BH}^3}\right)^{1/12}.
\label{eq:TXchemRDLrel}
\end{equation}

This estimate of temperature is consistent as long as $T_{\chi,\text{rel}}(t_\text{eq})\gtrsim m_\chi/3$, or equivalently as long as:
\begin{equation}
\beta\gtrsim\beta_\text{rel}\equiv
\frac{32\times2^{9/14}\pi^{159/14}}{54675\times3^{1/7}5^{5/14}\zeta^{20/7}(3)}\frac{g_{*,V}^{1/14}}{g_{*,\chi}^{4/7}}\frac{1}{\sqrt{\gamma}}\left(\frac{m_\chi^4M_\text{BH}^3}{M_\text{Pl}^7}\right)^{1/7}.
\label{eq:betarel}
\end{equation}

The ratio of the temperatures of two sectors is given by 
\begin{equation}
\xi=\frac{T_{\chi,\text{rel}}(t_\text{chem-eq})}{T_V(t_\text{chem-eq})}=2^{11/8}(5\pi)^{1/8}\frac{1}{g_{*,V}^{1/8}}\gamma^{1/8}\beta^{1/4}\left(\frac{M_\text{BH}}{M_\text{Pl}}\right)^{1/4}.
\label{eq:ratioTXtoTVatteqRDLrel}
\end{equation}
The efficiency of number-changing processes when DM particles become non-relativistic at $t=t_m$, can be estimated by comparing the rate of these processes with the Hubble expansion rate at $t=t_m$.
By using Eqs.~(\ref{eq:3to2ratetoHRel}), (\ref{eq:TVtauRD}), and (\ref{eq:energyinjectedtoDSRD2}) we have  
\begin{equation}
\frac{\Gamma_{\chi,3\rightarrow 2}(t_m)}{H(t_m)}\simeq\frac{5^{3/4}}{2^{5/4} e^6 \pi^{17/4}}\frac{g_{*,\chi}^2}{g_{*,V}^{3/4}}\gamma^{1/4}\sqrt{\beta}\frac{\sqrt{M_\text{BH}M_\text{Pl}}}{m_\chi}.
\label{eq:3to2rateRDLrelattm}
\end{equation}

If $\Gamma_{\chi,3\rightarrow 2}(t_m)/H(t_m)<1$, DM particles decouple at $t_m$ as soon as they become non-relativistic and temperature of the dark sector at $t_m$ sets the relic abundance of the dark sector today. This condition defines an upper bound on initial abundance of PBHs, $\beta_\text{RC}$, given by
\begin{equation}
  \beta_\text{RC}\equiv\frac{4\sqrt{2}\pi^{17/2}e^{12}}{5\sqrt{5}}\frac{g_{*,V}^{3/2}}{g_{*,\chi}^4}\frac{1}{\sqrt{\gamma}}\frac{m_\chi^2}{M_\text{BH}M_\text{Pl}},
    \label{eq:betacannrel}
\end{equation}
such that an initial abundance of PBHs in the range $\beta_\text{rel}\lesssim\beta\lesssim\beta_\text{RC}$ leads to a dark sector with a relativistic chemical equilibrium and no cannibal phase prior to decoupling.
Since from Eqs.~(\ref{eq:betarel}) and (\ref{eq:betacannrel}), we have $\beta_\text{rel}/\beta_\text{RC}\gg 1$, this scenario is not possible and therefore for a dark sector which is populated by Hawking evaporation of PBHs and is relativistic at equilibrium time, i.e.~ $\beta_\text{rel}\lesssim\beta\lesssim\beta_\text{crit}$, a cannibal phase is inevitable. The cannibal phase continues until $t_\text{dec}$, when the rate of number-changing processes becomes smaller than the Hubble expansion rate.
Following Eq.~(\ref{eq:TVtdecoupling}), $t_\text{dec}$ is given by 
\begin{eqnarray}
\nonumber T_V(t_\text{dec})&\simeq&
\frac{e^{3/2}\pi^{15/16}}{3\sqrt{3}\times 2^{9/16}5^{5/16}}\frac{g_{*,V}^{5/16}}{\sqrt{g_{*,\chi}}}\frac{1}{\gamma^{3/16}}\frac{1}{\beta^{3/8}}\left(\frac{m_\chi^{10}M_\text{Pl}}{ M_\text{BH}^3}\right)^{1/8}\\
&\times& W\left[\frac{3\times 5^{9/32}}{2\times 2^{15/32}e^{3/4}\pi^{51/32}}\frac{g_{*,\chi}^{3/4}}{g_{*,V}^{9/32}}\gamma^{3/32}\beta^{3/16}\left(\frac{M_\text{BH}M_\text{Pl}}{m_\chi^2}\right)^{3/16}\right]^{1/2}.
\label{eq:TVtdecRDcannLrel}
\end{eqnarray}
 Temperature of the dark sector at $t_\text{dec}$ which sets the relic abundance of the dark sector today is obtained from Eq.~(\ref{eq:TXdecrel}) as
\begin{equation}
 T_\chi (t_\text{dec})=\frac{2m_\chi}{W\left[6\,e^6\left(\frac{T_V(t_m)}{T_V(t_\text{dec})}\right)^6\right]}.
\label{eq:TXtdecRDcannLrel}
\end{equation}
After $t_\text{dec}$, DM particles decouple and the dark sector temperature scales with the expansion as the temperature of a decoupled non-relativistic gas while dark sector also develops a chemical potential as described by Eq.~(\ref{eq:TXafterdec}). 
The amount of DM today is given by Eq.~(\ref{eq:YXaftertdecRDcannLrel}) where $s(T)$, $T_V(t_\text{dec})$, and $T_\chi(t_\text{dec})$ are given by Eqs.~(\ref{eq:entropy}), (\ref{eq:TVtdecRDcannLrel}), and (\ref{eq:TXtdecRDcannLrel}) respectively.

\textbf{Chemical Equilibrium, Non-Relativistic Gas:} If the dark sector establishes chemical equilibrium when DM particles are non-relativistic, then by using Eqs.~(\ref{eq:TXchemnonrel}), (\ref{eq:TVtauRD}), (\ref{eq:avee}),  (\ref{eq:numberdensityinjectedtoDSRD}), and (\ref{eq:energyinjectedtoDSRD2}) the temperature of the dark sector at equilibrium is evaluated as
\begin{equation}
T_{\chi,\text{non-rel}}(t_\text{chem-eq})=\frac{2m_\chi}{3W\left[\frac{512\times 2^{4/9}\pi^{46/3}}{820125\times 3^{7/9}\zeta^{40/9}(3)}\frac{g_{*,V}^{1/9}}{g_{*,\chi}^{8/9}}\frac{1}{\gamma^{7/9}}\frac{1}{\beta^{14/9}}\left(\frac{m_\chi^8 M_\text{BH}^6}{M_\text{Pl}^{14}}\right)^{1/9}\right]}.
\label{eq:TXchemRDLnonrel} 
\end{equation}

This calculation of temperature is consistent as long as $T_{\chi,\text{non-rel}}(t_\text{eq})\lesssim m_\chi/3$, or equivalently as long as $\beta\lesssim\beta_\text{rel}$.

The ratio of temperatures of two sectors, for a fixed value of $\beta$, scales as
\begin{equation}
\xi=\frac{T_{\chi,\text{non-rel}}(t_\text{chem-eq})}{T_V(t_\text{chem-eq})}\sim\frac{m_\chi^{1/3}\sqrt{M_\text{BH}}}{M_\text{Pl}^{5/6}\text{ln}\,\frac{m_\chi^4M_\text{BH}^3}{M_\text{Pl}^7}}.
\label{eq:ratioTXchemRDLnonreltoTV}
\end{equation}
Therefore for fixed DM mass, the logarithmic dependence of dark sector temperature on PBHs mass in comparison with the power-law dependence of temperature of the visible sector on PBHs mass can lead to a cold dark sector (consists of non-relativistic particles) which is hotter than the visible sector.

The possibility of a cannibal phase can be examined by comparing the rate of number-changing processes with the Hubble expansion rate, which by using Eqs.~(\ref{eq:3to2ratetoHnonRel}), (\ref{eq:TVtauRD}), (\ref{eq:avee}),  (\ref{eq:numberdensityinjectedtoDSRD}), and (\ref{eq:energyinjectedtoDSRD2}) is estimated as
\begin{equation}
\frac{\Gamma_{\chi,3\rightarrow 2}(t_\text{chem-eq})}{H(t_\text{chem-eq})}\simeq
\frac{30267225703125\zeta^{10}(3)}{1048576\pi^{37}}\frac{g_{*,\chi}^4}{g_{*,V}}\gamma^2\beta^4\frac{M_\text{Pl}^4}{m_\chi^3M_\text{BH}}.
\label{eq:3to2rateattEQRDLnonrel}
\end{equation}

For those regions of parameter space that $\Gamma_{\chi,3\rightarrow 2}(t_\text{chem-eq})/H(t_\text{chem-eq})<1$, DM particles decouple and temperature of the dark sector at $t_\text{eq}$ sets the relic abundance of the dark sector. This condition defines a threshold for initial abundance of PBHs, $\beta_\text{NRC}$, given by
\begin{equation}
   \beta_\text{NRC}\equiv\frac{32\,\pi^{37/4}}{405\sqrt{3}\,5^{3/4}\zeta^{5/2}(3)}\frac{g_{*,V}^{1/4}}{g_{*,\chi}}\frac{1}{\sqrt{\gamma}}\left(\frac{m_\chi^3M_\text{BH}}{M_\text{Pl}^4}\right)^{1/4},
\label{eq:betacannnonrel}
\end{equation}
where an initial abundance of PBHs in the range $\beta_\text{kin}\lesssim\beta\lesssim\beta_\text{NRC}$, leads to a dark sector with a non-relativistic chemical equilibrium and no cannibal phase while an initial abundance of PBHs in the range $\beta_\text{NRC}\lesssim\beta\lesssim\beta_\text{rel}$, gives rise to a dark sector with a non-relativistic chemical equilibrium succeeded by a cannibal phase.

For $\beta_\text{kin}\lesssim\beta\lesssim\beta_\text{NRC}$, DM particles decouple at $t_\text{dec}=t_\text{chem-eq}$, and then temperature of the dark sector scales as the temperature of a decoupled non-relativistic gas while dark sector also develops a chemical potential as described by Eq.~(\ref{eq:TXafterdec}). 
The amount of DM today is given by Eq.~(\ref{eq:YXaftertdecRDcannLrel}) where $s(T)$, $T_V(t_\text{chem-eq})$, and $T_\chi(t_\text{chem-eq})$ are given by Eqs.~(\ref{eq:entropy}), (\ref{eq:TVtchemRD}), and (\ref{eq:TXchemRDLnonrel}) respectively.

For those regions of parameter space that $\Gamma_{\chi,3\rightarrow 2}(t_\text{chem-eq})/H(t_\text{chem-eq})\gtrsim1$, or equivalently,  $\beta_\text{NRC}\lesssim\beta\lesssim\beta_\text{rel}$, 
dark sector enters a cannibal phase which ends at $t_\text{dec}$. 
Temperature of the visible sector at $t_\text{dec}$ is obtained from Eq.~(\ref{eq:decouplingcondnonrelcann}) as
\begin{eqnarray}
\nonumber T_V(t_\text{dec})&\simeq&
\frac{2^{5/6}\pi^{5/2}}{3\times3^{1/3}\sqrt{5}\zeta^{5/6}(3)}\frac{g_{*,V}^{1/3}}{g_{*,\chi}^{2/3}}\frac{1}{\gamma^{1/3}}\frac{1}{\beta^{2/3}}\left(\frac{m_\chi^{11}}{M_\text{BH}^3M_\text{Pl}^2}\right)^{1/12}\sqrt{T_{\chi,\text{non-rel}}(t_\text{chem-eq})}\\
&\times&W\left[\frac{3^{11/12}5^{3/8}\zeta^{5/12}(3)}{4\times 2^{1/6}\pi^{19/8}}\frac{g_{*,\chi}^{5/6}}{g_{*,V}^{7/24}}\gamma^{1/6}\beta^{1/3}\left(\frac{M_\text{BH}^3M_\text{Pl}^8}{m_\chi^5T_{\chi,\text{non-rel}}^6(t_\text{chem-eq})}\right)^{1/24}\right]^{1/2},
\label{eq:TVtdecRDcannLnonrel}
\end{eqnarray}
and temperature of the dark sector at $t_\text{dec}$ is evaluated by using Eq.~(\ref{eq:TXdeccannibalnonrel}) as 
\begin{equation}
T_\chi (t_\text{dec})=\frac{2m_\chi}{W\left[\frac{1}{4\pi^3}g_{*,\chi}^2\left(\frac{T_V(t_\text{chem-eq})}{T_V(t_\text{dec})}\right)^6\frac{m_\chi^6}{s_\chi^2(t_\text{chem-eq})}\right]}. 
\label{eq:TXtdecRDcannLnonrel}
\end{equation}
After $t_\text{dec}$, DM particles decouple and temperature of the dark sector scales with the expansion as the temperature of a decoupled non-relativistic gas while dark sector also develops a chemical potential as described by Eq.~(\ref{eq:TXafterdec}). 
The amount of DM today is given by Eq.~(\ref{eq:YXaftertdecRDcannLrel}) where $s(T)$, $T_V(t_\text{dec})$, and $T_\chi(t_\text{dec})$ are given by Eqs.~(\ref{eq:entropy}), (\ref{eq:TVtdecRDcannLnonrel}), and (\ref{eq:TXtdecRDcannLnonrel}) respectively.

\subsubsection{\boldmath{$m_\chi > T_\textbf{\text{BH}}$}}
\textbf{Kinetic Equilibrium:} It is straightforward to show that immediately after evaporation, the rate of the elastic scattering processes is smaller than the Hubble expansion rate at that time and hence kinetic equilibrium cannot happen instantaneously. Following Eqs.~(\ref{eq:TVtauRD}), (\ref{eq:avee}) and (\ref{eq:numberdensityinjectedtoDSRD}), we have
\begin{equation}
\frac{\Gamma_{\chi, 2\rightarrow2}(\tau)}{H(\tau)}\sim\frac{2025\zeta^3(3)}{64\pi^{11}}g_{*,\chi}\frac{\beta}{\beta_\text{crit}}\left(\frac{T_\text{BH}}{m_\chi}\right)^4\lesssim\frac{2025\zeta^3(3)}{64\pi^{11}}g_{*,\chi}\simeq10^{-4}g_{*,\chi}.
\label{ratekinRDtauHrel}
\end{equation}
To check the possibility of establishing kinetic equilibrium afterwards, we estimate the maximum of the ratio of the rate of elastic scattering processes to the Hubble expansion  which corresponds to the time $t_m$, when the particles become non-relativistic. From Eqs.~(\ref{eq:tvattm}) and (\ref{eq:avee}), $t_m$ is given by
\begin{equation}
T_V(t_m)=\frac{15\zeta(3)}{\pi^4}T_V(\tau),
\label{eq:TVattmRDHrel}
\end{equation}  
and therefore we have
\begin{equation}
\frac{\Gamma_{\chi, 2\rightarrow2}(t_m)}{H(t_m)}\sim\frac{135\zeta^2(3)}{64\pi^7}g_{*,\chi}\frac{\beta}{\beta_\text{crit}}\left(\frac{T_\text{BH}}{m_\chi}\right)^4\lesssim\frac{135\zeta^2(3)}{64\pi^7}g_{*,\chi}\simeq10^{-3}g_{*,\chi}.
\label{ratekinRDtmHrel}
\end{equation}
It is clear that for radiation-dominated Universe, when $m_\chi>T_\text{BH}$, reaching kinetic equilibrium is not possible. This conclusion still holds even if self-coupling saturates its perturbative unitarity limit (see the discussion after Eq.~(\ref{ratekinRDtauLrel})).

\textbf{Chemical Equilibrium:} Since in a radiation-dominated Universe, when $m_\chi>T_\text{BH}$, it is not possible for the dark sector to reach kinetic equilibrium, establishing chemical equilibrium is definitely unattainable.

\subsection{Matter-dominated Universe \boldmath{$(\beta\geq \beta_\text{crit})$}}
If initial abundance of PBHs is large enough, since their energy density scales as the energy density of matter, they can eventually dominate the energy density and initiate an early matter domination epoch; after their evaporation, there will be a secondary reheating and subsequently a transition to a secondary radiation-dominated epoch. The SM particles produced by Hawking evaporation of PBHs, equilibrate quickly to the new reheating temperature of the Universe, $T_V(\tau)$, and the emitted DM particles evolve in this radiation-dominated Universe.

Immediately before evaporation of PBHs, the energy density of the Universe can be related to the lifetime of PBHs via Friedemann equation:
\begin{equation}
H^2(\tau)=\left(\frac{2}{3\tau}\right)^2=\frac{8\pi\rho_\text{BH}(\tau)}{3M_\text{Pl}^2},
\label{eq:friedmannMD}
\end{equation}

to obtain the energy density of the Universe stored in PBHs at the time of their evaporation, $\rho_\text{BH}(\tau)=M_\text{BH}n_\text{BH}(\tau)$, in terms of their lifetime: 
\begin{equation}
\rho_\text{BH}(\tau)=\frac{M_\text{Pl}^2}{6\pi\tau^2}.
\label{eq:BHenergydensityMD}
\end{equation}

The injected energy into the visible sector and dark sector, $\rho_{V,\text{BH}}(\tau)$ and $\rho_{\chi,\text{BH}}(\tau)$ respectively, can be written as
\begin{equation}
\rho_{V,\text{BH}}(\tau)=n_\text{BH}(\tau)\bar{E}_V\sum_{i\in\text{SM}} N_{V,i},~~~~~~\rho_{\chi,\text{BH}}(\tau)=n_\text{BH}(\tau)N_\chi\bar{E}_\chi
\label{eq:energyinjectedMD}
\end{equation}
The reheating temperature of the Universe, $T_V(\tau)$, provided an instantaneous equilibration, equals to:
\begin{equation}
T_V(\tau)=\frac{1}{32\sqrt{2}\times 5^{1/4}\pi^{5/4}}\frac{\sqrt{g_{*,V}+g_{*,\chi}}}{g_{*,V}^{1/4}}\left(\frac{M_\text{Pl}^5}{M_\text{BH}^{3}}\right)^{1/2}
\simeq\frac{1}{32\sqrt{2}\times 5^{1/4}\pi^{5/4}}g_{*,V}^{1/4}\left(\frac{M_\text{Pl}^5}{M_\text{BH}^{3}}\right)^{1/2}
\label{eq:TVtauMD}
\end{equation}

and the initial number density of DM particles is given by
\begin{equation}
n_\chi(\tau)=n_\text{BH}(\tau)N_\chi=\frac{N_\chi}{6\pi}\frac{M_\text{Pl}^2}{M_\text{BH}\tau^2}.
\label{eq:numberdensityinjectedtoDSMD}
\end{equation}

\subsubsection{\boldmath{$m_\chi < T_\textbf{\text{BH}}$}}
\textbf{Kinetic and Chemical Equilibrium:} It can be easily shown that in this case also, immediately after evaporation, the rate of the elastic scattering processes is smaller than the Hubble rate and therefore kinetic equilibration cannot happen instantaneously. Following Eqs.~(\ref{eq:TVtauRD}), (\ref{eq:avee}) and (\ref{eq:numberdensityinjectedtoDSMD}), we have
\begin{equation}
\frac{\Gamma_{\chi, 2\rightarrow2}(\tau)}{H(\tau)}\sim\frac{675\zeta^3(3)}{16\pi^{11}}g_{*,\chi}\simeq2.5 \times 10^{-4} g_{*,\chi}.
\label{ratekinMDtauLrel}
\end{equation}
A larger self-coupling, even as large as the perturbative unitarity limit, cannot lead to a fast kinetic equilibrium after evaporation (see the discussion after Eq.~(\ref{ratekinRDtauLrel})).
However, Kinetic equilibrium may be established later at $t=t_\text{kin-eq}$ which is given by
\begin{equation}
T_V(t_\text{kin-eq})\sim\frac{675\zeta^3(3)}{16\pi^{11}}g_{*,\chi}T_V(\tau),
\label{eq:ratekineticMDEQLrel}
\end{equation}
and this should happen before $t_m$, or equivalently for $T_V(t_\text{kin-eq})\gtrsim T_V(t_m)$, which by using $T_V(t_m)=T_V(\tau)m_\chi/\bar{E}_\chi$, leads to the following upper bound on the mass of PBHs:
\begin{equation}
    M_\text{BH}\lesssim M_\text{BH,kin}\equiv\frac{45\zeta^2(3)}{128\pi^8}g_{*,\chi}\frac{M_\text{Pl}^2}{m_\chi}.
     \label{eq:MBHkin}
\end{equation}

By using Eqs.~(\ref{eq:Tkinteq}), (\ref{eq:avee}), (\ref{eq:TVtauMD}), and (\ref{eq:numberdensityinjectedtoDSMD}) we can estimate the temperature of the dark sector after reaching kinetic equilibrium at $t_\text{kin-eq}$ as:
\begin{equation}
T_\chi(t_\text{kin-eq})=\frac{15\zeta^2(3)}{128\pi^{8}}g_{*,\chi}\frac{M_\text{Pl}^2}{M_\text{BH}}.
\label{eq:TkintMDLrel}
\end{equation}

Eq.~(\ref{eq:chemeqcondition}) can be used to find the time at which the dark sector reaches chemical equilibrium, $t_\text{chem-eq}$ as
\begin{equation}
T_V(t_\text{chem-eq})= 
\frac{45\times3^{1/3}5^{1/12}\zeta^{5/3}(3)}{16\times2^{5/6}\pi^{83/12}}g_{*,V}^{1/4}g_{*,\chi}^{1/3}\left(\frac{m_\chi^4 M_\text{Pl}^7}{M_\text{BH}^5}\right)^{1/6}.
\label{eq:TVtchemMD}
\end{equation}

\textbf{Chemical Equilibrium, Relativistic Gas:} If the dark sector attains chemical equilibrium while DM particles are still relativistic, the temperature of the dark sector can be obtained by using Eqs.~(\ref{eq:TXchemrel}), (\ref{eq:avee}), (\ref{eq:energyinjectedMD}), (\ref{eq:TVtauMD}), and (\ref{eq:numberdensityinjectedtoDSMD}) as:
\begin{equation}
T_{\chi,\text{rel}}(t_\text{chem-eq})=T_V(t_\text{chem-eq}),
\label{eq:TXchemMDLrel}
\end{equation}
which is equivalent to $\xi=1$. This may be understood by the fact that the injected energy into each sector is proportional to the numbers of degrees of freedom of that sector and the final temperature of each sector is also determined by the same number of degrees of freedom.

The estimated temperature of the dark sector is consistent as long as $T_{\chi,\text{rel}}(t_\text{chem-eq})\gtrsim m_\chi/3$, or equivalently as long as:
\begin{equation}
M_\text{BH}\lesssim M_\text{BH,rel}\equiv
\frac{405\times5^{3/10}\zeta^2(3)}{32\times2^{4/5}\pi^{83/10}}g_{*,V}^{3/10}g_{*,\chi}^{2/5}\left(\frac{M_\text{Pl}^7}{m_\chi^2}\right)^{1/5}.
\label{eq:MBHrel}
\end{equation}

The plausibility of a cannibal phase prior to decoupling can be determined by using Eqs.~(\ref{eq:3to2ratetoHRel}), (\ref{eq:energyinjectedMD}), and (\ref{eq:TVtauMD}) to estimate the ratio of rate of number-changing processes to the Hubble expansion rate as
\begin{equation}
\frac{\Gamma_{\chi,3\rightarrow 2}(t_m)}{H(t_m)}\simeq\frac{\sqrt{5}}{16e^6\pi^{9/2}}\frac{g_{*,\chi}^2}{\sqrt{g_{*,V}}}\frac{M_\text{Pl}}{m_\chi}.
\label{eq:3to2rateMDLrelattm}
\end{equation}

Since $\Gamma_{\chi,3\rightarrow 2}(t_m)/H(t_m)\gg1$, a cannibal phase is inevitable and the occurrence of an RNC thermal history is not possible. Therefore PBHs with masses less than $M_\text{BH,rel}$ lead to an RC thermal history. The cannibal phase continues until $t_\text{dec}$, when the rate of number-changing process becomes smaller than the Hubble expansion rate. 
Following Eqs.~(\ref{eq:TVtdecoupling}) and (\ref{eq:TXdecrel}), temperature of the visible sector and temperature of the dark sector at $t_\text{dec}$ which sets the relic abundance of the dark sector today are given by 
\begin{equation}
T_V(t_\text{dec})=\frac{2\sqrt{2}e^{3/2} \pi^{9/8}}{3\sqrt{3}\times 5^{1/8}}\frac{g_{*,V}^{1/8}}{\sqrt{g_{*,\chi}}}\left(\frac{m_\chi^5}{M_\text{Pl}}\right)^{1/4}W\left[\frac{3\times5^{3/16}}{4\sqrt{2}e^{3/4}\pi^{27/16}}\frac{g_{*,\chi}^{3/4}}{g_{*,V}^{3/16}}\left(\frac{M_\text{Pl}}{m_\chi}\right)^{3/8}\right]^{1/2},
\label{eq:TVtdecMDcannLrel}
\end{equation}
and, 
\begin{equation}
 T_\chi (t_\text{dec})= \frac{2m_\chi}{W\left[6\,e^6\left(\frac{T_V(t_m)}{T_V(t_\text{dec})}\right)^6\right]},
 \label{eq:TXtdecMDcannLrel}
\end{equation}
respectively. After $t_\text{dec}$, DM particles decouple and temperature and chemical potential of the dark sector evolve as described by Eq.~(\ref{eq:TXafterdec}). 
The amount of DM today is given by Eq.~(\ref{eq:YXaftertdecRDcannLrel}) where $s(T)$, $T_V(t_\text{dec})$, and $T_\chi(t_\text{dec})$ are given by Eqs.~(\ref{eq:entropy}), (\ref{eq:TVtdecMDcannLrel}), and (\ref{eq:TXtdecMDcannLrel}) respectively. It is worth mentioning that the amount of DM today only depends on DM mass, $m_\chi$ and is independent of the mass of PBHs, $M_\text{BH}$.

\textbf{Chemical Equilibrium, Non-Relativistic Gas:} If the dark sector reaches chemical equilibrium when DM particles are non-relativistic, then by using Eqs.~(\ref{eq:TXchemnonrel}), (\ref{eq:avee}), (\ref{eq:energyinjectedMD}), (\ref{eq:TVtauMD}), and (\ref{eq:numberdensityinjectedtoDSMD}) the temperature of the dark sector at equilibrium cab be expressed as

\begin{equation}
T_{\chi,\text{non-rel}}(t_\text{chem-eq})=\frac{2m_\chi}{3W\left[\frac{8192\times 2^{5/9}\pi^{145/9}}{18225\times3^{5/9}5^{2/9}\zeta^{40/9}(3)}\frac{1}{g_{*,\chi}^{8/9}g_{*,V}^{2/3}}\left(\frac{m_\chi^8M_\text{BH}^{20}}{M_\text{Pl}^{28}}\right)^{1/9}\right]}.
\label{eq:TXchemMDLnonrel} 
\end{equation}

This estimate of temperature is consistent as long as $T_{\chi,\text{non-rel}}(t_\text{chem-eq})\lesssim m_\chi/3$, or equivalently as long as $M_\text{BH}\gtrsim M_\text{BH,rel}$.

For a fixed PBHs initial abundance, $\beta$, the ratio of temperatures of two sectors scales as
\begin{equation}
\xi=\frac{T_{\chi,\text{non-rel}}(t_\text{chem-eq})}{T_V(t_\text{chem-eq})}\sim\frac{m_\chi^{1/3}M_\text{BH}^{5/6}}{M_\text{Pl}^{7/6}\text{ln}\,\frac{m_\chi^2M_\text{BH}^5}{M_\text{Pl}^7}},
\label{eq:ratioTXchemMDLnonreltoTV}
\end{equation}
which shows that for fixed DM mass, the logarithmic dependence of dark sector temperature on PBHs mass in comparison with the power-law dependence of temperature of the visible sector on PBHs mass can lead to a cold dark sector (consisting of non-relativistic particles) which is hotter than the visible sector.

To check the possibility of a cannibal phase, we need to estimate the rate of number-changing processes at equilibrium moment, $t_\text{eq}$, and compare it with Hubble expansion rate at the same time, which by using Eqs.~(\ref{eq:3to2ratetoHnonRel}), (\ref{eq:avee}), (\ref{eq:energyinjectedMD}), (\ref{eq:TVtauMD}), and (\ref{eq:numberdensityinjectedtoDSMD}) is evaluated as
\begin{equation}
\frac{\Gamma_{\chi,3\rightarrow 2}(t_\text{chem-eq})}{H(t_\text{chem-eq})}\simeq
\frac{1.16\zeta^{10}(3)}{\pi^{39}}g_{*,\chi}^4 g_{*,V}\frac{M_\text{Pl}^8}{m_\chi^3M_\text{BH}^5}.
\label{eq:3to2rateattEQMDLnonrel}
\end{equation}

For those regions of parameter space that $\Gamma_{\chi,3\rightarrow 2}(t_\text{chem-eq})/H(t_\text{chem-eq})<1$, DM particles decouple and temperature of the dark sector at $t_\text{eq}$ sets the relic abundance of the dark sector. This condition defines a threshold for mass of PBHs, $M_\text{BH,NRC}$, given by
\begin{equation}
     M_\text{BH,NRC}\equiv\frac{45\times3^{3/5}\zeta^2(3)}{64\times2^{2/5}\pi^{39/5}}g_{*,\chi}^{4/5}g_{*,V}^{1/5}\left(\frac{M_\text{Pl}^8}{m_\chi^3}\right)^{1/5},
    \label{eq:MBHcannnonrel}
\end{equation}
where PBHs with a mass in the range, $M_\text{BH,NRC}\lesssim M_\text{BH}\lesssim M_\text{BH,kin}$, leads to an NRNC thermal history while  in the range, $M_\text{BH,rel}\lesssim M_\text{BH}\lesssim M_\text{BH,NRC}$, gives rise to an NRC thermal history.

For $M_\text{BH,NRC}\lesssim M_\text{BH}\lesssim M_\text{BH,kin}$, $t_\text{dec}=t_\text{chem-eq}$, and After $t_\text{dec}$, DM particles decouple and temperature and chemical potential of the dark sector evolve as described by Eq.~(\ref{eq:TXafterdec}). 
The amount of DM today is given by Eq.~(\ref{eq:YXaftertdecRDcannLrel}) where $s(T)$, $T_V(t_\text{chem-eq})$, and  $T_\chi(t_\text{chem-eq})$ are given by Eqs.~(\ref{eq:entropy}), (\ref{eq:TVtchemMD}), and (\ref{eq:TXchemMDLnonrel}) respectively.

For those regions of parameter space that $\Gamma_{\chi,3\rightarrow 2}(t_\text{chem-eq})/H(t_\text{chem-eq})\gtrsim1$, or equivalently for $M_\text{BH,rel}\lesssim M_\text{BH}\lesssim M_\text{BH,NRC}$, 
dark sector enters a cannibal phase which ends at $t_\text{dec}$. 
Eqs.~(\ref{eq:decouplingcondnonrelcann}) and (\ref{eq:TXdeccannibalnonrel}) can be used to evaluate the temperature of the visible sector at $t_\text{dec}$ as
\begin{eqnarray}
\nonumber T_V(t_\text{dec})&\simeq& \frac{8\times2^{2/3}\pi^{17/6}}{3^{11/12}5^{1/6}\zeta^{5/6}(3)}\frac{1}{g_{*,\chi}^{2/3}}
\left(\frac{m_\chi^{11} M_\text{BH}^5}{M_\text{Pl}^{10}}\right)^{1/12}\sqrt{T_{\chi,\text{non-rel}}(t_\text{chem-eq})}\\
&\times&W\left[\frac{3^{17/24}5^{5/24}\zeta^{5/12}(3)}{8\times2^{7/12}\pi^{61/24}}\frac{g_{*,\chi}^{5/6}}{g_{*,V}^{1/8}}\left(\frac{M_\text{Pl}^{16}}{m_\chi^5 M_\text{BH}^5 T_{\chi,\text{non-rel}}^6(t_\text{chem-eq})}\right)^{1/24}\right]^{1/2}.
\label{eq:TVtdecMDcannLnonrel}
\end{eqnarray}
and temperature of the dark sector at $t_\text{dec}$ as
\begin{equation}
T_\chi (t_\text{dec})=\frac{2m_\chi}{W\left[\frac{1}{4\pi^3}g_{*,\chi}^2\left(\frac{T_V(t_\text{chem-eq})}{T_V(t_\text{dec})}\right)^6\frac{m_\chi^6}{s_\chi^2(t_\text{chem-eq})}\right]}.
\label{eq:TXtdecMDcannLnonrel}
\end{equation}

After $t_\text{dec}$, DM particles decouple and temperature and chemical potential of the dark sector evolve as described by Eq.~(\ref{eq:TXafterdec}). 
The amount of DM today is given by Eq.~(\ref{eq:YXaftertdecRDcannLrel})
where $s(T)$, $T_V(t_\text{dec})$, and $T_\chi(t_\text{dec})$, are given by Eqs.~(\ref{eq:entropy}), (\ref{eq:TVtdecMDcannLnonrel}), and  (\ref{eq:TXtdecMDcannLnonrel}) respectively.

\subsubsection{\boldmath{$m_\chi > T_\textbf{\text{BH}}$}}
\textbf{Kinetic Equilibrium:} Since the rate of the elastic scattering processes is smaller than the Hubble expansion rate at the evaporation time, hence kinetic equilibrium cannot happen instantaneously. This can be shown by using Eqs.~(\ref{eq:avee}), (\ref{eq:TVtauMD}) and (\ref{eq:numberdensityinjectedtoDSMD}), to estimate the ratio of the rate of elastic scattering processes to the Hubble expansion rate as:
\begin{equation}
\frac{\Gamma_{\chi, 2\rightarrow2}(\tau)}{H(\tau)}\sim\frac{675\zeta^3(3)}{16\pi^{11}}g_{*,\chi}\left(\frac{T_\text{BH}}{m_\chi}\right)^4\lesssim\frac{675\zeta^3(3)}{16\pi^{11}}g_{*,\chi}\simeq2.5\times 10^{-4}g_{*,\chi}.
\label{ratekinMDtauHrel}
\end{equation}

By estimating the maximum of the ratio of the rate of elastic scattering processes to the Hubble expansion rate which corresponds to the time $t_m$, when the particles become non-relativistic, we can examine the the possibility of occurrence of kinetic equilibrium afterwards. By using Eq.~(\ref{eq:TVattmRDHrel}), we have 
\begin{equation}
\frac{\Gamma_{\chi, 2\rightarrow2}(t_m)}{H(t_m)}\sim\frac{45\zeta^2(3)}{16\pi^7}g_{*,\chi}\left(\frac{T_\text{BH}}{m_\chi}\right)^4\lesssim\frac{45\zeta^2(3)}{16\pi^7}g_{*,\chi}\simeq10^{-3}g_{*,\chi}.
\label{ratekinMDtmHrel}
\end{equation}
we conclude that in a matter-dominated Universe, when $m_\chi>T_\text{BH}$, reaching kinetic equilibrium is nor feasible. Increasing the self-coupling cannot change this result (see the discussion after Eq.~(\ref{ratekinRDtauLrel})).

\textbf{Chemical Equilibrium:} In a matter-dominated Universe, when $m_\chi>T_\text{BH}$, it is not possible for the dark sector to reach chemical equilibrium, simply because kinetic equilibrium cannot be reached.


\end{document}